\documentstyle[a4,12pt,mydefs,psfig]{article}
\newcommand{\CUP}{\mathop{\cup}}
\newcommand{\be}{\begin{equation}}
\newcommand{\ee}{\end{equation}}
\newcommand{\la}{\left|\begin{array}{ccc}}
\newcommand{\ra}{\end{array}\right|}
\newcommand{\OR}{orthogonality}
\newcommand{\BE}{Biedenharn-Elliot relations}
\newcommand{\RA}{Racah identity}
\begin{document}

\begin{center}
{ \LARGE \bf Topological quantum field theory \\and invariants of
graphs\\[0.25cm]
for quantum groups}
 \\[1.5cm]

\begin{tabular}{cc}
Anna Beliakova \footnotemark[1]
& Bergfinnur Durhuus\\
 Institut f\"{u}r Theoretische Physik &   Mathematics Institute\\
 Freie Universit\"{a}t Berlin
& University of Copenhagen\\
\end{tabular}
\vspace*{0.5cm}
\\August 93 \vspace*{0.5cm}

\end{center}

\parbox{10cm}{{\footnotesize {\bf Abstract}: On  basis of
generalized 6j-symbols we give a formulation of topological quantum
field theories for 3-manifolds including observables in the form of
coloured graphs. It is shown that the 6j-symbols associated with
deformations of the classical groups at simple even roots of unity
provide examples of this construction.
Calculational methods are developed which, in particular, yield the
dimensions of the state spaces as well as a proof of the relation,
previously announced for the case of $SU_q(2)$ by V.Turaev,
 between these models and corresponding ones based on
the ribbon graph construction of Reshetikhin and Turaev.
}}

\section{Introduction}\label{ein}
\footnotetext[1]{supported by DAAD and DFG, SFB 288
''Differentialgeometrie und Quantenphysik``}
In ref. [TV] a novel combinatorial approach to 3-dimensional
topological quantum field theory was proposed. Its basis is the
observation that the 6j-symbols of $SU_q (2)$ obey the symmetries of a
tetrahedron and satisfy identities which may also be interpreted
geometrically in terms of glued tetrahedra and which lead to the
possibility of associating state sums (partition functions) with
3-dimensional triangulated manifolds which are independent of the
triangulation, i.e.\ they are topological invariants.

This approach was generalized in [DJN, D] to a large class of
algebras (replacing $SU_q(2)$) with associated generalized 6j-symbols
thus leading to a class of (unitary) 3-dimensional topological quantum
field theories satisfying all the standard properties (see [Wi],
[At]).

In the case of  $SU_q(2)$ a second generalization was introduced in
[KS] by including observables in the form of coloured graphs on the
boundary or the interiour of the manifolds. This leads to
effective calculational methods which were used to calculate  e.g.\
the dimensions of the state spaces in this model.

In this paper we combine the approaches of [DJN, D] and [KS] by
providing a simple geometric interpretation of the state sums
introduced in [KS]. This leads to a simplification of the discussion
in [KS] in a more general setting, including the case of 6j-symbols
associated to quantum deformations of an arbitrary classical group.

As an application we calculate the dimensions of the state spaces in
the general case in terms of the fusion matrices and we give a proof
that the partition function $Z(M)$ associated to a closed 3-manifold $M$ is
related to the invariant $\tau(M)$ introduced by Reshetikhin and
Turaev [RT2] in terms of ribbon graphs by
\be Z(M)=|\tau(M)|^2 \ee
for quantum deformations of the classical groups at the even simple
roots of unity.

The paper is organized as follows. In section 2 we formulate a general
system of axioms for 6j-symbols appropriate for our construction. In
section 3 we construct the state sums
   $Z(M,G_{\underline{x}})$ where   $M$ is a 3-manifold and $G$ is a graph
on the boundary $\partial M$ whose lines are coloured by labels
indicated by
${\underline{x}}$, and we discuss the geometric meaning of
$Z(M,G_{\underline{x}})$. Section 4 is devoted to an analysis of the
properties of $Z(M,G_{\underline{x}})$, as e.g.\ its behaviour under
cutting of handles or removal of tubes in $M$. This analysis yields
the desired calculational tools which are applied in section 5 to
evaluate the dimensions of the state spaces and to establish eq.\
(1.1). In section 6 we establish the properties of 6j-symbols stated
in section 2 in the case of a quantum group by ribbon graph
techniques. In addition, we prove that the state sum for a planar
graph coincides with the corresponding ribbon graph invariant of [RT1].

\section {Abstract 6j-symbols}\label{sixj}
In this section we list the defining properties of the abstract
6j-symbols to be used in the subsequent construction.

 Let  $I$ be a finite set with involution
$\ast : I\to I $ ($i\mapsto i^\ast$) and a distinguished element $0=0^\ast $.
The elements in $I$ will be called colours. To each triple of colours
 $(i,j,k)\in I^3$ there  is associated a finite dimensional complex
 vector space $V^k_{ij}$ of dimension $N^k_{ij}$, and we assume there
exist canonical isomorphisms

\be V^k_{ij}\simeq V^{i^\ast }_{j{k^\ast }}\, , \;\;\;
V^{k^\ast }_{j^\ast i^\ast }\simeq{(V^k_{ij})}^\ast \ee
where ${(V^k_{ij})}^\ast $ denotes the dual vector space of $V^k_{ij}$, and
\be V^k_{ij}\simeq V^k_{ji}\, .\ee
In the following these isomorphisms will be used without further
notice to identify these spaces. Moreover, we assume that
\be N^0_{ij^\ast}=\delta_{ij} \; .\ee
% This implies that:
%$$N^k_{ij}=N^k_{ji}=N^{i^\ast}_{j k^\ast}=N^{k^\ast}_{i^\ast
%j^\ast}.$$

In  [DJN] a rather general framework for the construction of
Hilbert spaces $V^k_{ij}$ fulfilling these properties was given, where
$I$ labels a set of irreducible representations of an algebra
satisfying certain properties. The isomorphism (2.2) (see [DJN,
section 5]) is, however, not needed for the construction of
a topological quantum field theory for manifolds without graphs, but
is crucial when graphs with vertices of order 4 are present, as will be
seen below.

The first isomorphism in (2.1) allows us to associate a vector space
with each coloured oriented 2-simplex (triangle) as follows. Let
$\sigma^2$ be an oriented 2-simplex with boundary links
$\sigma^1_1\, , \sigma^1_2\, , \sigma^1_3$ decorated by  arrows as
indicated in Fig.1,

\input epsf.sty
\begin{center}
\mbox{\epsfysize=2.5cm
\epsffile{drei.ps}}
\\Fig.1

\end{center}
where the orientation of $\sigma^2$ is supposed to be clockwise. A
colouring of the links in $\sigma^2$ is a mapping
$j: \sigma^1_i\mapsto j(\sigma^1_i)\in I,\, i=1,2,3$.
Given a colouring $j$ of $\sigma^2$ we associate with $(\sigma^2, j)$
the vector space
\be V(\sigma^2, j)=V^{j(\sigma^1_3)}_{j(\sigma^1_1)
j(\sigma^1_2)}\;.\ee

If an arrow on a link $\sigma^1_i$ is reversed we replace in this
definition $j(\sigma^1_i)$ by $j^\ast (\sigma^1_i)$. The first
isomorphism in (2.1) shows that this definition is
 invariant under rotations of the triangle
 in $R^2$ and the second shows that the vector space associated with
the same triangle with reversed orientation is the dual of the
original one.

Next, we assume the existence of a mapping
 $i\mapsto \omega_i$  from $I$ to ${\bf C}/ \{0\}$ such that
$ \omega_0=1,\omega^2_{k}=
\omega^2_{k^\ast}$ and
\be \label{eins}
 \sum_k \omega^2_k N^k_{ij}=\omega^2_i \omega^2_j \, ,
\;\;\;\; \sum_i {\omega_i}^4=\omega^2\neq 0\;.
\ee
In [DJN] $\omega^2_i$ was denoted by $F^{-1}_i$ and in the case of a
quantum group, to be discussed later, it equals the $q$-dimension
of the representation $i$ up to a sign. The assumption
$\omega^2_k=\omega^2_{k^\ast} $ then expresses the reality of the
$q$-dimension while the first relation in (2.5) reflects the
multiplicativity  of
the $q$-dimension
w.r.t.\ tensor products.
%\be \sum_{ij} \omega^2_i \omega^2_j N^k_{ij}= \omega^2 \omega^2_k \;.\ee

Finally, to each ordered
 6-tuple $(i,j,k,l,m,n)\in I^6$,
%such that $(i,j,k), (k,l,m), (i,n,m)$ and
%$(j,l,n)$  are admissible, , we assume there
is associated
an abstract 6j-symbol,
%which is the element of non-ordered
%tensor product over $K$ of four intertwiner spaces:
 \be  \label{zwei} \la
  i&j&k\\
  l&m&n
\ra \in V^k_{ij}\otimes V^m_{kl} \otimes V^n_{i^\ast m} \otimes
V^l_{j^\ast n} .\ee

By the association of vector spaces with coloured 2-simplexes
discussed above we see that the tensor product in (2.6) may be
associated to the boundary of the coloured tetrahedron depicted in
Fig.2.

\begin{center}
\mbox{\epsfysize=3cm
\epsffile{tet.ps}}
\\Fig.2

\end{center}
In order that the 6j-symbol in (2.6) define a unique vector
 associated to the tetrahedron it must be invariant under the
tetrahedral symmetry group, i.e.\

 \be  \la
i&j&k\\
l&m&n
\ra = \la
        k^\ast&i&j^\ast\\
        n&l&m
\ra = \la
                      l&m^\ast&k^\ast\\
                      i&j^\ast&n^\ast \ra . \ee

In addition, we shall assume that
\be \la
i&j&k\\
l&m&n
 \ra =  \la
              j&i&k\\
             m^\ast& l^\ast&n^\ast
\ra ,\ee
which is seen to imply that reversal of orientation of the tetrahedron
is equivalent to applying the involution $\ast$ to all labels.

Besides the symmetry relations (2.7-8) the 6j-symbols are assumed to
satisfy the following four relations, where products of 6j-symbols mean
(unordered) tensor products together with contraction w.r.t.\ mutually
dual pairs of vector spaces associated with certain pairs of factors:

1) Orthogonality:
\be \sum_k \omega^2_k \la i&j&k\\
                      C&B&A \ra{ \la i&j&k\\C&B&A^\prime \ra}^\ast =
\omega^{-2}
 _A \delta_{A A^\prime} 1_{V^B_{iA}\otimes V^A_{jc}}, \ee
where $${ \la i&j&k\\l&m&n\ra}^\ast = \la i^\ast&j^\ast&k^\ast\\
l^\ast&m^\ast& n^\ast\ra \, , $$
2)  \BE:
\be \sum_n \omega^2_n \la i&j&k\\l&m&n \ra \la i&n&m\\D&A&C \ra \la j&l&n\\
D&C&B \ra
= \la i&j&k\\ B&A&C \ra \la k&l&m\\ D&A&B \ra , \ee
3) Racah identities: There exists a mapping  $k\mapsto q_k$  from $I$ to ${\bf
C}/\{0\}$, such that  $q_0=1$, $q_k=q_{k^\ast}$ and
\be  \frac{q_k}{q_i q_j} \la i&j&k\\A&B&C \ra= \sum_D\omega^2_D  \frac{q_A
 q_B} {q_C q_D} \la i&A&D\\j&B&C \ra \la j&i&k\\ A&B&D \ra. \ee
Moreover, we assume that the replacement of all $q_i$ in (2.11) by $q^{-1}_i$
leads to another identity.

4) Considering $1_{V^B_{iA}}$ as a vector in
$V^B_{iA}\otimes(V^B_{iA})^\ast$ and identifying $V^i_{i0}$, $i\in I$,
with $\bf C$ according to (2.3) we have
 \be \la i^\prime&A^\prime&B\\ A^\ast&i&0 \ra = \frac{ \delta_{AA^\prime}
\delta_{ii^\prime}}{\omega_i \omega_A} 1_{V^B_{iA}}\; . \ee

The 6j-symbols defined in [DJN]  with the notation
\be F_{pq} \left[ \begin{array}{cc} i&j\\k&l\end{array} \right]=
\omega^2_p \la   i&j&q\\l&k&p\ra \ee
for the quantum groups obtained as deformations of the universal
enveloping algebra  $U_q {\sf g}$ of the classical semisimple Lie
algebra  $ {\sf g}$, were shown to satisfy the assumptions (2.7), (2.9) and
(2.10) if $q$ is an even simple root of unity and   $I$ is a certain
set of irreducible representations of  $U_q
{\sf g}$. These were shown to
be sufficient for  the
construction of a topological quantum field theory for manifolds without
graphs.
 The additional relations (2.8), (2.12) and Racah identities will turn out
to be of importance for the inclusion of graphs with vertices of oder
4. These relations were not discussed in [DJN], but they are relatively easy
to establish in the case of quantum groups by the methods developed
there, with
$$q^2_i=i(c)\; $$
for each representation $i\in I$, where $c$ is a certain central
element in the quantum group (acting as multiplication by a scalar
$q^2_i$, since $i$ is irreducible). In fact,
it is  straight forward to check (2.12) and it is
 possible to see that (2.8) follows from the Racah identities.

As to the latter one defines the braiding operator
$$ B_{pq} \left[ \begin{array}{cc} i&j\\k&l\end{array} \right]:
V^k_{jp}\otimes V^p_{jl}\to V^k_{jq}\otimes V^q_{il}$$
by setting its matrix element to be given as
\be < B_{pq} \left[ \begin{array}{cc} i&j\\k&l\end{array} \right]
\alpha\otimes \beta, \delta\otimes \gamma>=\alpha\circ \beta\circ
\check{R}^{ij}_{12} \circ \gamma^\ast\circ \delta^\ast \ee
for $\alpha\in V^k_{jp}, \beta\in V^p_{jl}, \delta\in V^k_{jq},
\gamma\in V^q_{il}$, where the RHS is an intertwiner between $k$ and
$k$ and hence a number, and, if $V_i$ denotes the representation
space of $i$, $\check{R}^{ji}_{12}:V_j\otimes V_i\to
V_i\otimes V_j$ is given by (see [DJN],[D])
\be \check{R}^{ji}_{12}= \sigma\circ(j\otimes i)(R)\, ,\ee
where $\sigma(x\otimes y)=y\otimes x$ for $x\in V_j,y\in V_i$ and
$R\in  U_q {\sf g}\otimes  U_q {\sf g}$
is the standard $R$-matrix.

Defining the isomorphism
 $V^k_{ij}\simeq V^k_{ji}$ by
\be \beta\mapsto\hat{\beta}=\frac{q_k}{q_i q_j} \beta\circ
\check{R}^{ij}_{12}
\ee
we have $\beta=\hat{\hat{\beta}} $ and one finds that
\be \omega^{-2}_p B_{pq} \left[ \begin{array}{cc} i&j\\k&l\end{array} \right]=
\frac{q_p q_q}{q_k q_l}\la
i&l&q\\j&k&p\ra \ee
which is the vector that will be associated to a 4-vertex (or  rectangle) (see
Fig.6 below).

Using the definition (2.14) it is quite easy to verify eq. (2.11) by
using complete reducibility of tensor products of representations in
$I$ and the properties of $R$ (eq. (5.4) in [DJN]). Replacing in (2.16)
$q_i\to q^{-1}_i$ and $R\to R^{-1}$ in  a similar way we obtain the
second Racah identity.

In section 6 we describe an alternative way of establishing the
properties of the 6j-symbols.

\section{ State sums for  manifold with coloured graphs}
 \label{fun}
\newtheorem{satz}{Theorem}[section]
\newtheorem{lem}[satz]{Lemma}
Let be $M$ an oriented compact 3-manifold with
triangulation $X$ whose 1-simplexes (links) are assumed to be
decorated by arrows in some arbitrary way, and let
 $\underline{j}:X\ni \sigma^1 \mapsto j(\sigma^1)\in I$ be a colouring of
  1-simplexes in $X$. Furthermore,
let $G$ be an oriented graph on the boundary $\partial M$, i.e.\ a
finite  1-dimensional
simplicial complex without boundary, which is compatible with $X$, i.e.\ the
1-skeleton of $G$ is contained in the
 1-skeleton of $\partial X$. We assume that
 $G$  has only 2-, 3- and 4-vertices.
The  4-vertices can be of two
types:
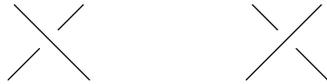
\begin{center}
\mbox{\epsfysize=1cm \epsffile{vver.ps}}\end{center}
 which we shall call inverse to each other.  Orientations of links
 on  opposite sides of 2- and
 4-vertices are assumed to  coincide. A colouring of $G$ is a mapping
  ${\underline{x}}$ from its lines (maximal connected sets of links joined by
vertices of order 2) such that the colours on opposite sides of
4-vertices are identical. The graph $G$ coloured by
 ${\underline{x}}$ will be denoted by  $G_{\underline{x}}$.

We associate now with $(M,X,G_{\underline{x}})$ a new triangulated
pseudo-manifold $(M_{G_{\underline{x}}},
 X_{G_{\underline{x}}})$ as follows: Let  $ {\cal C}=\{
c_1,\ldots,c_n\}$ denote the set of connected components of
 $\partial M /G$,
which we shall identify with new additional vertices. For each triangle
$\sigma^2\in c_i$ we glue onto $\partial M$ a new tetrahedron, which has base
$\sigma^2$ and an opposite vertex $c_i$. Furthermore, we glue these
tetrahedra together along triangles, which they share, i.e.\ for two
triangles
 $\sigma^2_1\in c_i$
and $\sigma^2_2\in c_i$ with $  \sigma^2_1\cap \sigma^2_2=[AB]$
the corresponding tetrahedra are glued along the triangle
$(ABc_i)$ (see Fig.3).
\begin{center}
\mbox{\epsfysize=4cm
\epsffile{rdr.ps}  }
\\ Fig.3
\end{center}
Next, for each
 $k$-coloured link $\sigma^1\in \partial X$,
contained in  an $x$-coloured line of
$G_{\underline{x}}$, and which is contained in the boundaries of the
 components $c_i, c_j$ in $\cal C$  (Fig. 4),
we glue on a tetrahedron along
the two triangles containing $\sigma^1$ and $c_i$, respectively $c_j$,
and continually glue along common triangles of the added tetrahedra.

\begin{center}
\mbox{\epsfysize=4cm
\epsffile{rlin.ps}}
\\ Fig.4
\end{center}

To the so obtained simplicial complex we glue for each vertex $v$ of
order 3 a new tetrahedron which contains the vertices
  $(v,
c_i, c_j, c_k)$, where $c_i, c_j, c_k$ are the  components in
$\cal C$   sharing  $v$ (Fig.5).

\begin{center}
\mbox{\epsfysize=4cm
\epsffile{gab.ps}}
\\ Fig.5
\end{center}
This finishes the construction of the coloured triangulated
pseudo-manifold
 $(M_{G_{\underline{x}}},
 X_{G_{\underline{x}}})$. The corresponding uncoloured triangulated
pseudo-manifold will be denoted  $(M_{G},
X_{G})$. Its principal feature is that it contains $(M,X)$
in its interiour except for the original 4-vertices in $G$, which are
now 4-vertices in $\partial X_G$. It is convenient to view the four
triangles in $\partial X_G$ sharing such a 4-vertex $v$ as making up
a rectangle with $v$ as center. Then $\partial X_G$ consists of a set
of triangles and rectangles, namely one triangle for each 3-vertex in
$G$ and one rectangle for each 4-vertex in $G$, and hence $\partial
X_G$  may be viewed as a dual graph to $G$ in $\partial M$.

We next associate  with
 $(M,{G_{\underline{x}}})$ the state sum
\be Z(M,{G_{\underline{x}}})=Z(M_{G_{\underline{x}}}),\ee
where $ Z(M_{G_{\underline{x}}})$ is
defined in analogy with [TV, DJN] as follows: Given a colouring
$\underline{j}$ of $X$, a factor
 $\omega^{-2}$ is attached  to each interiour vertex $\sigma^0\in
intX_{G_{\underline{x}}}$,  a factor
$\omega^2_i$ to each
 $i$-coloured interiour link $\sigma^1\in intX_{G_{\underline{x}}}$
and a
6j-symbol to each tetrahedron
 $\sigma^3\in X_{G_{\underline{x}}}$ as described above. Finally, to
each coloured rectangle in $\partial X_{G_{\underline{x}}}$, as
depicted in Fig.6,

\begin{center}
\mbox{\epsfysize=4cm
\epsffile{recht.ps}}
\\ Fig.6
\end{center}
 whose orientation is assumed to be clockwise, we
attach the braiding operator
\be \label{r}\frac{q_A q_B}{q_C q_D} \la y&A&D\\ x&B&C \ra \in
V^D_{Ay}\otimes V^B_{Dx} \otimes V^C_{By^\ast} \otimes V^A_{Cx^\ast}\,. \ee

For the  inverse 4-vertex the   braiding operator can be obtained
from (3.2) by replacement of all $q_i$ by $q^{-1}_i$.
% If the $y$-colouring line lie above the $x$-colouring one,
%then in
%(\ref{r})  all $q_i$ would be replace   with $q^{-1}_i$, what corresponds
% ${\cal R}^{-1}$.
%We now associate and  ${\cal R}$-Matrix
%(resp.\ ${\cal R}^{-1}$) to each rectangle.
We then form the product of all factors so associated to
 $ X_{G_{\underline{x}}}$, contract w.r.t.\ all interiour triangles as
well as the triangles inside the rectangles in  $\partial
X_{G_{\underline{x}}}$ and take the direct sum over all  colourings
$\underline{j}$ of $X$, thus obtaining the desired vector

\be Z(M_{G_{\underline{x}}})\in V(\partial M, G_{\underline{x}})\, ,\ee
where $ V(\partial M, G_{\underline{x}})$ is the tensor product of
vector spaces associated to the coloured triangles in $ \partial
M_{G_{\underline{x}}}$, which are dual to 3-vertices in $
{G_{\underline{x}}}$.
Note that  $Z(M_{G_{\underline{x}}})$ is independent of the initial
distribution of arrows on the links in $X$, since $i\mapsto i^\ast$ is
a bijective mapping on $I$.

This definition of $ Z(M_{G_{\underline{x}}})$
is identical to the one in [DJN] except for the issues concerning
4-vertices in $G$ and that no factors $\omega^{-1}$ are associated to
vertices in $ \partial X_{G_{\underline{x}}}$. By the same arguments
as in [DJN] it then follows that $ Z( M_{G_{\underline{x}}})$ {\it is
independent of the interiour  of the triangulation $  X_{G_{\underline{x}}}$}
as a consequence of the identities (2.9-10). In particular,
it is independent of the triangulation $X$ of $M$ and is invariant
under homotopy changes of $ G_{\underline{x}}$, since the latter
clearly only affect the interiour of  $ X_{G_{\underline{x}}}$.

In general $M_G$ is not a manifold. For example, in case $G$ is empty
one point $c_i$ will be added to $X$ for each connected component of
$\partial M$ and the points corresponding to components not
homeomorphic to $S^2$ will be non-manifolds points of $M_G$.
On the other hand, if $G$ is sufficiently
''large''  then $M_G$ will be
homeomorphic to $M$. A particular graph of this type on a connected
surface $\Sigma$ of genus $g\ge 1$ may be obtain by first choosing a
homology basis $a_1, ..., a_g, b_1, ..., b_g$ whose representatives
have one point $P$ in common and then deforming it such that $P$ is
separated into $4g-2$ vertices of order 3. We shall adopt the notation
of [KS] and call this graph $G^\Sigma$. The dual graph to $G^\Sigma$
 on $\Sigma$  is easily seen to yield a proper triangulation\footnotemark[2]
 of
$\Sigma$ and $M_{G^\Sigma}$ is obtained from $M$, $\partial M=\Sigma$,
by gluing on a cylinder $\Sigma\times [0,1]$ as is easy to verify.
\footnotetext[2]{ We use here a more general notion of  triangulation
than the standard one,
in that we allow identifications of subsimplexes of co-dimension $ > 1$
in a simplex of maximal dimension.}

In general, we have for a graph $G$ {\it without vertices of order 4} such
that $M_G$ is homeomorphic to $M$ that the partition function
$Z^\prime(M)$ of [DJN] is given by
\be Z^\prime(M)=\omega^{-N_G}\otimes_{\underline{x}}
Z(M,G_{\underline{x}})\; , \ee
where $N_G$ is the number of connected components of $\partial M/G$
and the triangulation of $\partial M$ is given by the dual graph to
$G$. In particular,
\be Z^\prime(M)\in V(\partial M, G) \, ,\ee
where
\be V(\partial M, G)\equiv \oplus_{\underline{x}} V(\partial M,
G_{\underline{x}} ) \ee
is the vector space associated to the surface $\partial M$
triangulated by the dual graph to $G$.

Given a closed, compact, oriented surface $\Sigma$, a graph $G$ in
$\Sigma$ and a compact oriented 3-manifold $M$ with $\partial
M=\Sigma$, we call $G$ a {\it canonical graph} in $\Sigma$ if $M_G$ is
homeomorphic to $M$, and we note that this property only depends on
$\Sigma$ and $G$ but not on $M$. In particular, a graph $G$ on
$\Sigma$ with no vertices of order 4 is canonical if and only if its
dual graph in $\Sigma$ yields a triangulation of $\Sigma$.

The correspondence (3.4) allows us to rephase results of [DJN] in
terms of manifolds with graphs. Of particular interest is the gluing
property (axiom (3) of [At]) whose triangulated version holds by
construction for $Z^\prime(M)$ (see [DJN]) and may be reformulated
as follows.

\begin{satz} Let $M$ be a (connected or disconnected) oriented compact
3-manifold with $\partial M=\Sigma_1\cup \Sigma_2\cup \Sigma \cup
\Sigma^\ast$ (disjoint union), where
$\Sigma_1 , \Sigma_2 , \Sigma ,
\Sigma^\ast$ are closed oriented surfaces such that there exists an
orientation reversing diffeomorphism $F:\Sigma\to \Sigma^\ast$.
Moreover, let $G^1_{\underline{x_1}}\subseteq \Sigma_1$ and
$G^2_{\underline{x_2}}\subseteq \Sigma_2$
be arbitrary coloured graphs and let $G\subseteq \Sigma$ be a
canonical graph without vertices of order 4 and $G^F$ its image in
$\Sigma^\ast$ by $F$. Then, if $M_\Sigma$ denotes the 3-manifold
obtained by gluing $M$ along $\Sigma$ and $\Sigma^\ast$ by $F$, we have
\be Z(M_\Sigma, G^1_{\underline{x_1}}\cup G^2_{\underline{x_2}})= \omega^{-2
N_G} \sum_{\underline{x}} \prod_{l\subseteq G} \omega^2_{x_l}(
 Z(M, G^1_{\underline{x_1}}\cup G^2_{\underline{x_2}}\cup
G_{\underline{x}}\cup G^F_{\underline{x}}))_{G_{\underline{x}}} \, ,
\ee
where the product on the RHS is over lines $l$ in $G$ and $(\cdot
)_{G_{\underline{x}}}$ indicates a contraction w.r.t.\ all pairs
of vector spaces associated to coloured 3-vertices in $G_{\underline{x}}$
and $G^F_{\underline{x}}$.
\end {satz}

%These two statements follow from the independence of
% $Z(M_{G_{\underline{x}}})$ on the interior triangulation of
%$M_{G_{\underline{x}}}$, which was proved in [DJN]. The proof
%given in [KMS] can be easy adapted to this case too.
% $\hfill \Box$

This theorem generalizes Thm. 7.1 of [KS] and is identical to eq.
(4.6) in [DJN] in case $G^1$ and $G^2$ are canonical graphs, but its
validity for arbitrary $G^1$ and $G^2$ follows, as in [DJN],
immediately from the construction of $Z(M_G)$ and its independence of
the interiour of the triangulation  $X_G$ of $M_G$.

It should be noted that  in eq. (3.4) we have not exhibited the
dependence of $Z^\prime(M)\equiv Z^\prime(M,G)$ on $G$ for the
following reason. Let $G$ be as in eq. (3.4), i.e.\ $G$ is a canonical
graph in $\partial M \equiv \Sigma$ without 4-vertices, and consider
the cylindrical 3-manifold $\Sigma\times [0,1]$  oriented such that we
may identify $\Sigma^\ast$ with $\Sigma\times \{0\}$ and $\Sigma $
with  $\Sigma\times \{1\}$. Then
$$P_{(\Sigma,G)}\equiv Z^\prime(\Sigma\times[0,1],  G\times \{0\},
 G\times \{1\})\in {V(\Sigma, G)}^\ast \otimes V(\Sigma, G)\simeq Hom
(V(\Sigma, G), V(\Sigma, G))\, .$$
Since $ M\cup_\Sigma
(\Sigma\times[0,1]) $ is homeomorphic to $M$ it follows from eq. (3.7)
that
$$Z^\prime(M,G)=P_{(\Sigma, G)} Z^\prime(M,G)\in P_{(\Sigma, G)}V(\Sigma,G)
\equiv V^\prime (\Sigma, G)\, .$$
Here we note that factors $\omega^2_x$ in eq. (3.7) associated to the
coloured lines in $G_{\underline{x}}$ are included in the definition
of the bilinear pairing of $V(\Sigma^\ast,G)$ and $V(\Sigma, G)$ (see
eq. (3.30) in [DJN]).

It follows, moreover, from eq. (3.7) (see section 4 of [DJN] for
details) that $P_{(\Sigma, G)}$
is a projection  and that $V^\prime (\Sigma, G_1)$ and $V^\prime
(\Sigma, G_2)$ for two arbitrary canonical graphs in $\Sigma$ may be
consistently identified by the mappings
$$P_{(\Sigma,G_1),(\Sigma, G_2)}\equiv Z^\prime(\Sigma\times[0,1],  G_1\times
\{0\},
 G_2\times \{1\})\in {V(\Sigma, G_1)}^\ast \otimes V(\Sigma,
G_2)\simeq $$
$$Hom
(V(\Sigma, G_1), V(\Sigma, G_2))\, $$
by which the vectors $Z(M,G_1)$ and $Z(M,G_2)$ are also identified.

In force of these identifications we thus obtain the vector space
$$V_\Sigma\simeq V^\prime (\Sigma, G)\subseteq V(\Sigma, G)$$
containing the vector
$$Z^\prime(M)\equiv Z^\prime(M,G)$$
for all 3-manifolds $M$ with $\partial M=\Sigma $.

By eq. (3.7) and the fact that $P_{(\Sigma, G)}$ is a projection we
have
\be dim V_\Sigma= tr  P_{(\Sigma, G)} =Z^\prime (\Sigma\times
S^1)=Z(\Sigma\times S^1) \, .\ee

For later use we note the following simple lemma, valid for any
compact oriented 3-manifolds $M,M_1,M_2$, and where we use the notation
$$Z(M)\equiv Z(M,{\emptyset})\, ,$$
with $\emptyset$ denoting the empty graph.

\begin{lem} i) If $D^3\subseteq int M$ is diffeomorphic to the 3-ball,
then, for any graph $G\subseteq \partial M$,
\be   Z(M, G)= \omega^{-2} Z(M/D^3, G)\, . \ee

ii) \be Z(S^3)= \omega^{-2},\;\;\; Z(D^3)=1\, .\ee

iii) If $M^\ast$ denotes $M$ with opposite  orientation, then
   \be \;Z(M)=Z(M^\ast)\, .\ee

iv) If $M$ is the connected sum of $M_1$ and $M_2$, i.e.\
 $M=(M_1/D^3_1)\CUP_{S^2} (M_2/D^3_2)$, where $D^3_1\subseteq int M_1$
and $D^3_2\subseteq int M_2$
 are diffeomorphic to the 3-ball and whose boundaries $S^2$
are identified, then
 \be Z(M, G_1\cup G_2)=\omega^{2} Z(M_1, G_1)\otimes
Z(M_2, G_2) \ee
for arbitrary graphs $G_1\subseteq \partial M_1$ and $G_2\subseteq
 \partial M_2$.
\end{lem}

{\bf Proof:}

{\em i)} Follows since $M_G=(M/D^3)_G$, but no factor $\omega^{-2}$
is associated to the vertex $c$ corresponding to the component
$\partial D^3 \subseteq \partial (M/D^3)$ in the definition  of
$Z(M/D^3,G)$.

{\em ii)} By triangulating $S^3$ by two tetrahedra the first relation
follows from (2.9) and the definition of $\omega^2$. The second
relation then follows from {\em i)} applied to $M=S^3$.

{\em iii)} Follows from (2.8) since all colours are summed over and
$i\mapsto i^\ast$ is a bijective mapping from $I$ to $I$.

{\em iv)} We first note that $M$ is diffeomorphic to
\be (M_1/D^3_1)\CUP_{S^2\times \{0\} } (S^2\times
[0,1])\CUP_{S^2\times \{1\}}(M_2/D^3_2) \, .\ee
Next we note that the graph $\overline{G}$ in $S^2$ consisting of two
3-vertices connected by three lines is canonical since its dual yields
a triangulation of $S^2$ by two triangles. We can obtain a
triangulation of $S^2\times [0,1]$ such that $S^2\times \{0\}$ and $
S^2\times \{1\}$ are both triangulated by two triangles by taking two
prisms, each triangulated in a standard fashion by three tetrahedra,
and gluing then together along their sides. Using this triangulation a
straightforward calculation using (2.9) and (2.5) yields
\be Z(S^2\times [0,1],  \overline{G}_{\underline{x}}\times\{0\},
 \overline{G}_{\underline{y}}\times\{1\} ) =\omega^2 1_{V^{x_3}_{x_1
x_2}} \otimes 1_{V^{y_3}_{y_1y_2}} \, ,\ee
where $\underline{x}=(x_1,x_2,x_3)$ and  $\underline{y}=(y_1,y_2,y_3)$
are the colours of the lines (suitably oriented)
in $ \overline{G}\times\{0\}$ and $ \overline{G}\times\{1\}$,
respectively. Applying eq. (3.7) to (3.13) the result follows from (3.14).
 $\hfill \Box$
\vspace*{0.3cm}

We close this section with some remarks on a few concepts that will be
of importance in the following (see also [KS]).

An {\it interiour graph} {\cal G}  in a 3-manifold $M$ consists of 1)
a core $c({\cal G}) $ which is a finite unoriented graph embedded in
$int M$, 2) a tubular neighbourhood ${\cal T_G }$ of $c({\cal G} )$ in
$int M$ and 3) an oriented graph $G$ in $\partial {\cal T_G }$
as introduced at the beginning of this section. If $G$ is coloured by
$\underline{x}$ we say that ${\cal G}$ is coloured by $\underline{x}$
and denote the coloured interiour graph by ${\cal G}_{\underline{x}}$.
A priori there need not be any connection between $c(\cal G)$ and $G$,
although this will be the case in the applications below. As a
particular example we mention the case in which each component $k_i$
of $c(\cal G)$ is homeomorphic to the circle $S^1$ and the corresponding
component ${\cal T}_i$ of $\cal T_G$ contains exactly one component
$l_i$ of $G$ and $l_i$ is homotopic to $k_i$ in ${\cal T}_i$. In this
case ${\cal G}$ is a framed link in $M$, and will usually be denoted by
$\cal L$.

For a compact oriented 3-manifold $M$ with an interiour coloured graph
${\cal G}_{\underline{x}}$  as above we define
 \be Z(M, {\cal G}_{\underline{x}})\equiv Z(M/{\cal T_G},
G_{\underline{x}}) \, . \ee
Of course, this definition can be generalized to the case where in
addition  a
graph is present in $\partial M$.

Of particular relevance for the following is the concept of a meridian
and of left- and righthanded lines. Assume that the 3-manifold $M$
contains an empty tube $T=S^1\times [0,1]\subseteq \partial M$, i.e.\
$M$ can be obtained from a 3-manifold $\tilde{M}$ by removing a
cylinder $C$ diffeomorphic to $D^2\times [0,1]$, where a $D^2$ is the
two-dimensional unit disc. A {\it meridian} on $T$ is an oriented
circle $m=S^1\times \{p\},p\in ]0,1[$.

A 3-manifold $M$ containing  a set of empty tubes $T_m,
T^\prime_{m^\prime}, ...$ with meridians $m, m^\prime, ...$ will be
denoted by $M(T_m, T^\prime_{m^\prime}, ...)$. Given a graph $G$ on
$\partial M$ the lines in $G$ may over- or undercross the meridians
$m, m^\prime, ...$, such that these together with $G$ constitute a
graph $G\cup m\cup m^\prime \cup ...\;\,$. For each colouring
$\underline{x}$ of $G$ we may thus define [KS]

\be Z(M(T_m, T^\prime_{m^\prime},
...),G_{\underline{x}})=\sum_{a,a^\prime, ...}
\frac{\omega^2_a}{\omega^2} \frac{\omega^2_{a^\prime}}{\omega^2} ...
Z(M,G_{\underline{x}}\cup m_a\cup m^\prime_{a^\prime}\cup ...\, )\,
,\ee
where the sum is over colourings of the meridians $m, m^\prime, ...\;\,$.

If a line $l$ in $G$ intersects a meridian $m$ exactly once we say
that $l$ is left-,  respectively right-, handed w.r.t.\ $m$ if $l$ over-,
respectively under-, crosses $m$. For the particular case of a framed link
${\cal L}$ as defined above the corresponding {\it left-, respectively
right-, handed link ${\cal L}_L$, resp. ${\cal L}_R$}, is obtained by
introducing a meridian $m_i$ on each of the tubes ${\cal T}_i$ such
that each component $l_i$ is left-, respectively \hbox{right-,} handed w.r.t.\
$m_i$.
The state sum associated to this left-, respectively \hbox{right-,} handed
framed link
is then given by (3.16) with the ${\cal T}_i$ replacing $T_m,
T^\prime_{m^\prime}, ... $ and will be denoted by $Z(M,{\cal L}_L)$,
resp. $Z(M,{\cal L}_R)$.

The main reason for introducing the definition (3.16) is illustrated
by the following lemma, which asserts that filling an empty tube is
equivalent (up to the factor $\omega^2$) to  introducing a meridian
on the tube.

\begin{lem} Let $\tilde{M}$ be a compact oriented  3-manifold that
contains  a cylinder $C=D^2\times [0,1]$,
with  $D^2 \times \{0,1\}\subseteq \partial \tilde{M}$, and let $M$
denote the manifold obtained by removing $C$ from $\tilde{M}$. Then
\be Z(\tilde{M}, G)=\omega^2 Z(M(T_m),G) \ee
for any graph $G\subseteq \partial \tilde{M}$  such that $G\cap
(D^2\times\{0,1\})  =\emptyset $ and where $T_m$ denotes the tube
$\partial D^2\times [0,1]\subseteq \partial M$ with meridian $m$.
\end{lem}
Pictorially (3.17) may be written as
\begin{center}
\mbox{\epsfysize=2.6cm \epsffile{tube.ps}}\end{center}
{\bf Proof:} This follows simply by realizing that the cylinder $C$
gets reinserted  automatically when $M_{G\cup m}$ is constructed. More
explicitly, for a given triangulation $X$ of $M$,
 $M_{G_{\underline{x}}\cup m_a}$ equals  $\tilde{M}_{G_{\underline{x}}}$
with a triangulation in which the (interiour) link connecting the
vertices $c_1$ and $c_2$ associated with the components in $\partial
M/(G\cup m)$ on either side of $m$ has fixed colour $a$, as indicated
on the following figure.
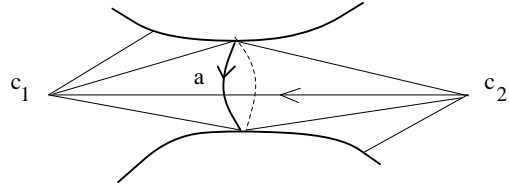
\begin{center}
\mbox{\epsfysize=2.4cm \epsffile{mer1.ps}}\end{center}
Thus inserting the factor $\omega^2_a$ and summing over $a$ one
obtains $Z(\tilde{M}_{G_{\underline{x}}})$ according to the definition
of the latter. $\hfill\Box$
\vspace*{0.3cm}

On the other hand the following result shows that a handle in $M$ can
be cut without changing the state sum.

\begin{lem} If the  3-manifold  $M$ contains a cylinder
$C=D^2\times[0,1]$ with
$\partial D^2\times [0,1] \subseteq \partial M$, and $\tilde{M}$
denotes the manifold obtained be removing $C$ from $M$, then
\be Z(M,G)=Z(\tilde{M},G) \ee
for any graph $G\subseteq \partial M$ such that
$G\cap(\partial D^2\times [0,1])=\emptyset$.
\end{lem}

{\bf Proof:} By choosing a triangulation $X$ of $M$ such that the
handle $C$ can be cut along a single triangle $t$ (with $\partial
t\subseteq \partial D^2 \times ]0,1[$), the result follows from eq.
(3.14) by an argument essentially identical to the one used in the
proof of Lemma 3.2
 $\hfill \Box$
\vspace{0.3cm}

Whereas Lemma 3.4 deals with cutting of handles disjoint from $G$ the
following lemma can be applied to handles traversed by a single line in
$G$.

\begin{lem} Let $M$ be a compact oriented 3-manifold and
$G_{\underline{x}}$ a coloured graph in $\partial M$.
If an   $x$-coloured line $L$ in ${G_{\underline{x}}}$ does not separate
two different components in  $\partial M/{G}$
and   there exists in
 $\partial M/{G}$ a contractible  loop in $M$
intersecting $L$ (transversally) only one, then
 \be Z(M,{G_{\underline{x}}})=0 ,\;\;\; when \;x\not=0 \;\; .\ee\end{lem}
{\bf Proof:}
Let $X$ be a triangulation of $M$ and denote by $A$ the vertex in
$X_G$ corresponding to the component of $\partial M/G$ containing $L$
in its boundary. Then there is an $x$-coloured link $l$ in $X_G$ (dual
to $L$) whose end-points both equal $A$.
The tetrahedron in $X_{G_{\underline{x}}}$ containing $l$ and a
$a$-coloured link in $L$ as opposite links then
 looks as follows:

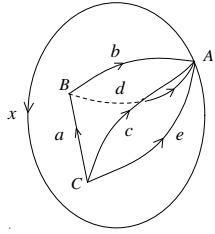
\begin{center}
\mbox{\epsfysize=3cm \epsffile{lin1.ps}}\end{center}
 By the assumption the two
triangles $(ABC)$ may be connected by a thickened disc. By
retriangulation of this thickened disk one may reach a (singular)
triangulation, in which the two triangles
 $(ABC)$ are identical. The link  $(BC)$ is then only contained in one
tetrahedron. Summing over its colour $a$ one obtains a contribution
%$Z(M,{G_{\underline{x}}})$
 $$\sum_a \omega^2_a \la a&b&c\\x&c&b \ra \;\; ,$$
contracted w.r.t.\ the dual pair of spaces $V^c_{ab}\, ,V^b_{a^\ast c}\,$.
But this expression equals $0$ for $x\not= 0$, since it may be
rewritten by eqs. (2.12) and (2.9)
as:
$$
%\sum_a \omega^2_a \hspace*{3.5cm}\vbox to 1.5cm{
%\special{psfile=lin2.ps hscale=40 vscale=40 voffset=-70 hoffset=-95}}
%=
\sum_a \omega^2_a \la a&b&c\\x&c&b\ra
\la a&b&c\\0&c&b\ra = \delta_{x0}1_{V^c_{c0}\otimes V^b_{b0}}\; .\;\;\hfill\Box
$$
\vspace*{0.3cm}

{\bf Remark 3.6} In Lemma 4.2 we shall see that a line with colour $0$
can be deleted from $G$ without changing $Z(M,G)$. Together with
Lemmas 3.4-5 this implies that if a single $x$-coloured line traverses
a handle, then the state sum is either zero (if $x\not= 0$)
or (if $x=0$) the line can be deleted and the handle cut.

\section{Calculational methods }\label{rech}
In this section we establish a number of invariance properties or
transformation rules for the state sums $Z(M,G)$ under local changes
of the graph $G$. These are generalizations of the results derived in
[KS], but we give different and, we hope, more transparent proofs.

In the following let
 $M$ be a compact oriented 3-manifold and
$G_{\underline{x}}$ a coloured graph in $\partial M$.

\begin{lem} If $G_{\underline{x}}=G^1_{\underline{x_1}}\cup
G^2_{\underline{x_2}} $, where $G^1$ is contained in a disc
$D^2\subseteq \partial M$ such that $G^2\cap D^2=\emptyset$, i.e.\ $G$
contains an isolated planar subgraph $G^1$, then
  \be \label{fak}
Z(M,{G_{\underline{x}}})=Z(G^1_{\underline{x_1}})\otimes
Z(M,{G^2_{\underline{x_2}}}), \;\ee
where
\be Z({G_{\underline{x}}})\equiv Z(D^3,{G_{\underline{x}}}) \ee
and $D^3$ denotes a 3-ball.
\end{lem}
{\bf Proof:}

By choosing a suitable triangulation $X$ of $M$ we can represent
  $(M,{G_{\underline{x}}})$
 as two manifolds
 $(M^\prime, G^1_{\underline{x_1}})$ and $(D^3,{G^2_{\underline{x_2}}})$,
 which are
glued together along the triangle $t\subseteq \partial M^\prime \cap
\partial D^3$ with $\partial t\subseteq \partial M$, and such that
$M^\prime$ is homeomorphic to $M$. Then $M_G$ equals the
  connected sum of
  $D^3_{G^1}$ and  $M^\prime_{G^2}$, obtained  by
first removing
 from  $D^3_{G^1}$, resp.\ from $M^\prime_{G^2}$, the tetrahedron
 with base $t$ and  opposite   vertex $c$, resp.\ $c^\prime$,
corresponding to the connected component of
 $\partial D^3/{G^1}$, resp.
$\partial M^\prime/ G^2$, containing $t$,
and subsequently gluing the two resulting manifolds  together along
the two copies of
$S^2$ (the boundaries of the removed tetrahedra), such that $c$ and
 $c^\prime$ are
identified.
Eq. (4.1) then follows from eq. (3.14) as in the proof of Lemma 3.2
{\it iv)}, taking into account that no factors $\omega^{-2}$ are
associated to   $c\in D^3_{G^1}$ and $c^\prime\in M^\prime_{G^2}$
nor to the vertex resulting from their identification in $M_G$.
$\hfill \Box$
\vspace*{0.3cm}

{\bf Examples:} 1) For a coloured circle $S^1_x$ we have

\be
\mbox{
\parbox{2.2cm}{$Z(S^1_x)=Z($}\parbox{1cm}{\hbox{\psfig{figure=kr.ps,height=1cm,width=0.8cm}}}
\parbox{1.3cm}{$)=Z($}\parbox{1cm}{\hbox{\psfig{figure=krei.ps,height=1cm,width=0.8cm}}}
\parbox{1.3cm}{$)=Z($}\parbox{1cm}{
\hbox{\psfig{figure=kreis.ps,height=1cm,width=0.8cm}}}
\parbox{1.4cm}{$)=\omega^2_x \, $.}
}
 \ee
Observe that, for a given triangulation $X$ of $D^3$, $\,(D^3_{S^1},
X_{S^1})$ equals $(S^3, X_{S^1})$ where $X_{S^1}$ contains two
distinguished vertices $c$ and $c^\prime$ connected by a link with
colour $x$. Triangulating $S^3$ by two tetrahedra eq. (4.3) follows
from eq. (2.9).

2)
 \be
%Z(\hspace*{2cm}\vbox to 2cm{\special{psfile=n.ps hscale=50
%vscale=50 hoffset=-53 voffset=-75}})
\mbox{
\parbox{0.7cm}{$Z\Biggl($ }
\parbox{1.9cm}{\hbox{\psfig{figure=n.ps,height=1.4cm,width=1.8cm}}}
\parbox{1.8cm}{$\Biggr)\; = 1_{V^k_{ij}} \, $.}
}\ee
Note that $D^3_G$, for the  graph in (4.4), equals a 3-ball with
boundary triangulated by two triangles and thus can be obtained by
gluing two tetrahedra along three common triangles. Eq. (4.4) is then
again a consequence of eq. (2.9).

3)
\be \mbox{
\parbox{0.7cm}{$Z\Biggl($
}\parbox{2.7cm}{\hbox{\psfig{figure=gabel.ps,height=2.3cm,width=2.7cm}}}
\parbox{4.2cm}{$\Biggr)\;\;
 = \la i&j&k\\
l&m&n \ra \, $.}   } \ee
For the graph in (4.5) $D^3_G$ equals the 3-ball whose boundary is
triangulated by the dual graph to $G$ and hence can be obtained as a
single tetrahedron. Eq. (4.5) then follows from the definition of
$Z(M,G)$.

4)
\be \mbox{
\parbox{0.7cm}{$Z\Biggl($
}\parbox{2.7cm}{\hbox{\psfig{figure=r.ps,height=2.6cm,width=2.7cm}}}
\parbox{4.3cm}{$\Biggr)\;
 =\frac{q_l q_m}{q_k q_n} \la
j&l&n\\ i&m&k \ra \,
$.}   } \ee
We leave the verification of this identity to the reader.

\begin{lem} The state sum $Z(M,G)$ is invariant under the following local
changes of
 ${G_{\underline{x}}}$:

i)
%\be \mbox{\epsfysize=3cm\epsffile{loop.ps}}\ee
\be \hbox{\psfig{figure=loop.ps,height=3cm}} \ee

ii)
%\be \mbox{\epsfysize=3cm\epsffile{orth.ps}}\ee
\be \hbox{\psfig{figure=orth.ps,height=3cm}} \ee
where the curly line indicates the contraction with respect to the dual pair
of spaces associated to the 3-vertices it connects.

iii)
%\be \mbox{\epsfysize=3cm\epsffile{null.ps}}\ee
\be \hbox{\psfig{figure=null.ps,height=3cm}} \ee
where the vector spaces $V^0_{x x^\ast}$ and $V^0_{y y^\ast}$ have
been identified with $\bf C$.
\end {lem}
{\bf Proof:}

{\it i)} The substitution (4.7) corresponds to the following local
transformation of the triangulation on the boundary of $M_G$:

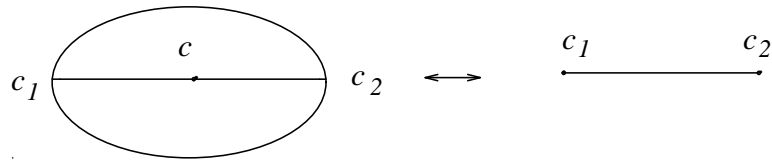
 \begin{center}
\mbox{\epsfysize=2cm \epsffile{loop1.ps}}\end{center}
where $c_1, c$ and $c_2$ are the vertices associated to the left,
middle  and the right components of $\partial M/G$, respectively on the
lefthand side of (4.7), and similarly for the righthand side.

Choosing a  triangulation of $M_{G_{\underline{x}}}$ which in the
vicinity of $c_1,c, c_2$ looks as

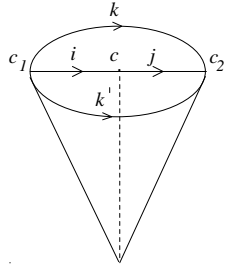
\begin{center}
\mbox{\epsfysize=3.5cm \epsffile{loop2.ps}} \end{center}
the result follows from
the orthogonality relations, which can be
interpreted in terms of a collapsing of two tetrahedra glued along two
common triangles
[DJN]:

\begin{center}
\mbox{\epsfysize=3.5cm \epsffile{loop3.ps}}\end{center}

{\it ii)} The substitution (4.8) corresponds to the following local
transformation of the triangulation on the boundary  $\partial
M_{G_{\underline{x}}}$:
\begin{center}
\mbox{\epsfysize=2cm \epsffile{orth1.ps}}\end{center}
 Choosing  a triangulation  of
 $ M_{G}$ which in the vicinity of $c_1,c_2, c_3, c_4$ looks as
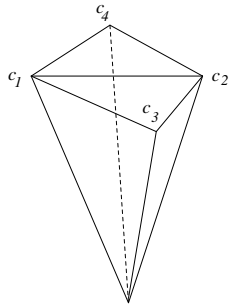
\begin{center}
\mbox{\epsfysize=4cm \epsffile{orth2.ps}}\end{center}
the contraction on the lefthand side of (4.8) corresponds to gluing the
two triangles $(c_1 c_2 c_4)$ and $(c_1 c_2 c_3)$ together. Having
done so we may again apply eq. (2.9) as above and obtain:
 \begin{center}
\mbox{\epsfysize=3.5cm \epsffile{orth3.ps}}\end{center}

which yields the desired result.

{\it iii)} The substitution (4.9) corresponds to the following local
transformation of the triangulation of the boundary of $M_G$:
\begin{center}
\mbox{\epsfysize=3cm \epsffile{null1.ps}}\end{center}
By choosing a suitable triangulation of $M_G$ one finds that (4.9) is
a simple consequence of eq. (2.12).
 $\hfill \Box$
\vspace*{0.3cm}

Next we note that the \RA\ (2.11) implies invariance of $Z(M,G)$ under
the following local substitution in $G$:

\be\mbox{\epsfysize=3cm \epsffile{rac.ps}}\ee
 which is easily seen in terms of a suitable local choice of
triangulation of $M_G$. Similarly, the  second Racah identity
 yields invariance under the substitution
\be
\mbox{\epsfysize=3cm \epsffile{qrac.ps}}\ee
By combining these  two identities with (4.8) one obtains invariance
of $Z(M,G)$ under the following local change of $G$:

\be
\mbox{\epsfysize=2.5cm \epsffile{duga.ps}} \ee
The \BE\ give rise to invariance under the substitution
\be
\mbox{\epsfysize=3.5cm \epsffile{be.ps}}\ee
which follows in a similar way as above by a convenient choice of
triangulation of $M_G$.

Combining (4.13) with (4.8) and (4.10) we get in addition invariance
under the substitution
\be
\mbox{\epsfysize=3cm \epsffile{yb.ps}}\ee
Of course, invariance of $Z(M,G)$ also holds under the substitution
analogous to (4.12) (resp. (4.13)), where the $j$-line (resp.
$a$-line) overcrosses the $i$-line (resp. $i$- and $j$-lines) on the
lefthand side.

In order to formulate the following lemma, which will be of importance
in the next section, we define the matrix $S$ by (see [KS])
\be S_{ab}=\omega^{-1} Z(\hspace*{1,8cm}
\includegraphics{s.ps}
) =\omega^{-1} \sum_c \frac{\omega^2_c
q^2_c}{q^2_a q^2_b} N^c_{ab}\; \ee
and set
\be \omega^2_b(a)=\frac{\omega}{\omega^2_a} S_{ab}\;.\ee
The last expression in eq. (4.15) follows by  first applying (4.8)
to the double line in the middle of the graph in (4.15), then applying
(4.10) and finally (4.4). We note that
\be  S_{ab}=S_{ba}=S_{a^\ast b^\ast}\; ,\;\;\ee
since $q_{a^\ast}=q_a, \omega_{c^\ast}=\omega_c$ and
$N^c_{ab}=N^c_{ba}=N^{c^\ast}_{a^\ast b^\ast}$, and that
\be \omega^2_b(0)=\omega S_{0b}= \omega^2_b \ee
since $q_0=\omega_0=1$ and $N^c_{0b}=\delta_{cb}$.

\begin{lem} i) The state sum $Z(M,G_{\underline{x}})$ is invariant
under the following local substitution in $G_{\underline{x}}$
 \be
\mbox{\epsfysize=3cm \epsffile{om.ps}} \ee

ii) We have
 \be \sum_b N^b_{cd} \omega^2_b(a)=\omega^2_c(a) \omega^2_d(a),\ee
i.e.\ $\omega^2_c(a), a\in I$, is an eigenvalue of the matrix $N_c$
defined by
\be (N_c)^b_d=N^b_{cd} ,\;\;\; b,d\in I\; ,\ee
for each $c\in I$, with eigenvector $(\omega^2_d(a))_{d\in I}$
(provided the latter is non-zero).

iii) If  $q^2_a\not= 1$ for all $a\in I/\{0\}$  and
\be \triangle=\sum_c q^2_c\omega^4_c
\not= 0,\ee
 then the following formulae hold:  \be \sum_b
\omega^2_b \omega^2_b (a)= \omega^2  \delta_{a0}, \ee
\be \sum_bS_{ab}S_{bc^\ast}=\delta_{ac}\, ,\ee
\be N^c_{ab}=\sum_d\frac{S_{ad}S_{bd}S_{c^\ast d}}{S_{0d}}\;\; .\ee
\end{lem}
{\bf Proof:}

The statements follow by rather obvious modifications of the arguments
in [KS], Appendix A. For completeness we give some details:

{\it i)} Using (4.8), Lemma 3.5, (4.9) and Lemma 4.1 we obtain
invariance under the following substitutions:

 %\begin{center}
%\special{psfile=s.ps hscale=40 vscale=40 voffset=-20 hoffset=-48}
%\mbox{ \epsfxsize=14.5cm \epsffile{om2.ps}}
\centerline{\hbox{\psfig{figure=om2.ps,height=2.6cm,width=14.3cm}}}
%\end{center}
as desired.

{\it ii)} Follows from {\it i)} and invariance under the following
substitutions:

%\begin{center}
%\mbox{\epsfysize=3.5cm \epsfxsize=14.5cm \epsffile{beweis2.ps}}
%\input{beweis.tex}
%\end{center}
\centerline{\hbox{\psfig{figure=beweis1.ps,height=3.5cm,width=14.3cm}}}

{\it iii)} It is easy to see (Lemma A.2 in [KS]) that the assumption
 $q^2_a\not=1$ and $\Delta \not= 0$ imply the existence of a
   $b\in I$, such that $\omega^2_b\not= \omega^2_b(a)
$. Then (4.23) follows for $a\not= 0$ from
$$(\omega^2_b - \omega^2_b(a) )\sum_c\omega^2_c\omega^2_{c^\ast}(a)=\sum_{cd}
(N^d_{bc}\omega^2_d\omega^2_{c^\ast}(a)- N^d_{bc^\ast}\omega^2_d(a)\omega^2_c)
=0 \, ,$$
where we have used {\it ii)} and the fact that $N^d_{b c^\ast}=N^c_{b d^\ast}$
by (2.1-2) as well as $\omega^2_{c^\ast}=\omega^2_c$. For $a=0$ eq.
(4.23) follows from the definition of $\omega^2$.
Eq. (4.24) coincides with eq. (4.23) for $a=0$ or $c=0$ in view of
eqs. (4.16-18). Multiplying by $\omega^2_b$ and summing over $b\in
I$ in (4.19) we get from eq. (4.23) invariance under the  substitution

\be
\mbox{\epsfysize=3cm \epsffile{omeg.ps}} \ee
Using this eq. (4.24) follows easily for  general $a,b\in I$ as in
[KS], Lemma A.3.

Finally, eq. (4.25) follows immediately from eqs. (4.20) and (4.16)
using (4.24).
$ \hfill \Box $
\vspace*{0.3cm}

We note that, in view of eq. (4.24), eq. (4.25) states that the matrix
$S$ {\it diagonalizes all the matrices $N_a; a\in I$}

 We end this section by discussing some useful properties of the
state sums for manifolds with interiour graphs, which can be derived
easily from the rules developed in this and the previous sections.
\vspace*{0.2cm}

1) Suppose that two meridians $m$ and $m^\prime$ are introduced on an
empty tube $T$ in $M$ disjoint from $G$. According to Lemma 3.3 this
is equivalent up to a factor $\omega^4 $ to filling the tube by two
cylinders in the vicinity of $m$ and $m^\prime$. Between the cylinders
there will then be an empty 3-ball. Filling this 3-ball is equivalent
to multiplying the state sum by $\omega^{-2}$ by Lemma 3.2 {\it i)}.
Thus, it follows that the presence of two  (or more) meridians on $T$
is equivalent to the presence of just one, i.e.\
\be Z(M(T_m,T_{m^\prime}),G)=Z(M(T_m),G). \ee
In fact, the purpose of the factors $\omega^{-2}$ in (3.16) is to
ensure this projection property of the meridians.

2) Consider in $M$ a branching of an empty tube $T_1$ into two empty
tubes $T_2$ and $T_3$, which are all disjoint from $G$ (see Fig.7).

\begin{center}
\mbox{\epsfysize=3cm \epsffile{branch.ps}}
\\Fig.7
 \end{center}
An argument similar to the previous one implies equivalence of any two
configurations of meridians on the tubes $T_1, T_2, T_3$, if only at
least two of the tubes contain a meridian.

3) Similarly one sees that lines in $G$ along a tube containing a
meridian may be deformed non-trivially as follows:

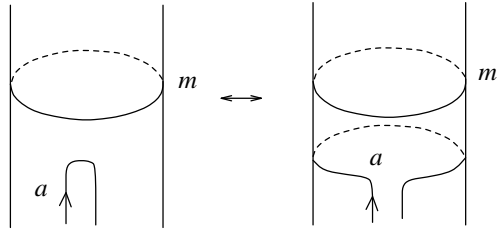
\begin{center}
\mbox{\epsfysize=3cm \epsffile{proj.ps}}  \end{center}

4) The statements in 1-3) may generalized to the case where lines in
$G$ (in addition to the $a$-line in 3) ) are traversing the tubes, if
only all these lines overcross (or all undercross) the meridians (and
the $a$-line in 3) ). This follows that by repeated use of (4.8) and
(4.12- 4.14).

5) Combining the generalized versions of 2) and 3) with Lemma 3.3 it
is easy to show that if $T_m$ and $T^\prime _{m^\prime}$ are two empty
tubes  in $M$ such that $T$, resp. $T^\prime $, is traversed by a
lefthanded , resp. righthanded, line then $T_m$ and $T^\prime _{m^\prime}$
have trivial braiding, i.e.\ they may be moved through each other.

6) Using (4.26) and 3) as well as Lemma 3.4 it follows that if a tube
in $M$ is traversed by an $a$-coloured line in $G$  which overcrosses
one meridian and undercrosses another then $Z(M,G)$ vanishes unless
$a=0$, in which case the line and the meridian can be deleted and the
tube filled:

\be
\mbox{\epsfysize=3cm \epsffile{twomer.ps}} \ee
We leave the details of these arguments to the reader. Alternatively,
[KS] can be consulted.
\vspace*{0.2cm}

Finally, we note the following lemma, which we shall need in the next section.
\begin{lem} Let ${\cal L}$ and  ${\cal L}^\prime$ be arbitrary  disjoint framed
links in $S^3$. Then the following holds.

{\it i)}
\be Z(S^3, {\cal L}_L\cup {\cal L}^\prime_R)= \omega^2 Z(S^3, {\cal L}_L)
Z(S^3, {\cal L}^\prime_R) \, .\ee

{\it ii)} For any colouring $\underline{a}$ of $\cal L$ we have
\be Z(S^3,({\cal L}_{\underline{a}})_L)=\omega^{-2}Z(G_L({\cal
L})_{\underline{a}})\;\; ,\ee
where $G_L({\cal L})$ is the planar graph (with no 3-vertices)
naturally obtained from the framed link $\cal L$ by projecting onto a
plane (and specified more precisely below).

{\it iii)} For any colouring $\underline{a}$ of $\cal L$ we have
\be Z(S^3,({\cal L}_{\underline{a}})_R)=\omega^{-2}Z(G_R({\cal
L})_{\underline{a}})\;\; ,\ee
where $G_R({\cal L})$ is obtained from $G_L({\cal L})$ by replacing
all 4-vertices by their inverses.
\end{lem}

{\bf Proof:}

{\it i)} follows immediately from 5) above and lemma 3.2 {\it iv)}.

{\it ii)} Consider $\cal L$ as embedded into ${\bf R}^3\subseteq S^3$
and choose a plane $\pi$ such that $\cal L$ lies on one side of $\pi$.
We may then deform $\cal L$ such that the circles on the tori
constituting the boundary of the tubular neighbourhood ${\cal T}_{\cal L}$
can be obtained as translates of the corresponding cores in the
direction orthogonal to and away from $\pi$. (In terms of ribbons
given by the framing of $\cal L$ this is equivalent to deforming the
ribbons such that no twists are present (see [RT1]).)
The coloured graph $G_L({\cal  L})_{\underline{a}}$ is then obtained
by projecting the circles on $\partial {\cal T}_{\cal L}$ onto $\pi$
taking  into account over- and undercrossings as well as orientation
and colouring $\underline{a}$ of $\cal  L$.

In order to prove (4.30) we then consider the crossings of tubes in
$\cal L$ (see the LHS of the figure below) corresponding to 4-vertices
in $G_L(\cal L)$. By Lemma 3.3 and 2) above we may connect the two
tubes at each crossing by a tube at the cost of a factor $\omega^2$.
Using the lefthandedness and the general version of 3) above (or
(4.8)) it follows that the two tubes may be joined as indicated in the
following figure.

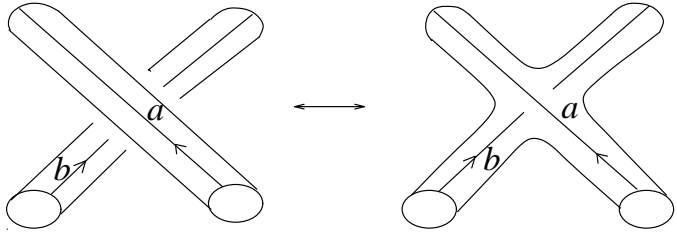
\begin{center}
\mbox{\epsfysize=3cm \epsffile{cros.ps}}
 \end{center}

One thus obtains an interiour graph in $S^3$ whose core is a copy of
$G_L({\cal L})$ and whose corresponding coloured graph  on the
boundary of its tubular neighborhood is a copy of $G_L({\cal
L})_{\underline{a}} $ and all of whose lines are lefthanded.

Now it is not difficult to see that all meridians can be eliminated,
partly by use of 2) above and partly by using lemmas
3.5 and 4.2 at the cost of factors $\omega^{-2}$.
Finally, handles (corresponding to the faces of the graph $G_L({\cal
L}) $) may be cut by lemma 3.4 thus obtaining $D^3$. Keeping track of
the $\omega$-factors one arrives at (4.30).

{\it iii)} Follows from {\it ii)} by observing that in the case of
righthanded lines the 4-vertex on the RHS of the figure above must be
replaced by its inverse. $\hfill\Box$

\section{Dimension of the state spaces and factorization of state
sums}\label{tqft}

This section is devoted to calculating the dimensions of the state
space $V_\Sigma$ associated to a closed, compact and oriented surface
$\Sigma$, and to showing that the state sum $Z(M)$ for a closed,
compact, oriented 3-manifold $M$ factorizes into a lefthanded and a
righthanded contribution, which in the  quantum group case equal $\tau
(M)$ and $\tau(M^\ast)$, respectively, where $\tau(M)$ is the
invariant introduced in [RT2].

Let us first note that by (3.8), (3.14) and (2.5) we have, for
$\Sigma$ connected of genus 0,
\be dim V_{S^2}=Z(S^2\times S^1)=\omega^{-4} \sum_{a,b,c} N^c_{ab}
\omega^2_a  \omega^2_b  \omega^2_c =1 \, . \ee
Let the matrix $|\vec{N}|^2$ be defined by
\be |\vec{N}|^2=\sum_a N^t_a N_a= \sum_a N_{a^\ast} N_a \ee
where the upper index $t$ denotes transposition and the last equality
follows from (2.1-2).

\begin{satz} Assume that  $q^2_a\not= 1$ for $a\not= 0$ and that $\Delta
\not= 0$.
Let $\Sigma$ be a connected, closed, compact, oriented
surface of genus $g\ge 1$. Then
\be dim V_{\Sigma}= (tr (|\vec{N}|^{2(g-1)}))^2 \, .\ee
\end{satz}
{\bf Proof:}

Given the computational rules developed in sections 3 and 4 the proof
of (5.3) is identical to the proof of the Thm. 7.4 (without punctures)
in [KS] with only minor modifications. It suffices therefore here to
indicate the main line of argument: According to eq. (3.8) we have to
calculate $Z(\Sigma\times S^1)$. Using Thm. 3.1 we may cut
$\Sigma\times S^1$ along $\Sigma\times \{a\}$, for some $a\in S^1$,
thus obtaining a manifold homeomorphic to $\Sigma\times [0,1]$ with a
graph $(G\times \{0\})\cup (G\times \{1\})$ on its boundary, where $G$
is a canonical graph on $\Sigma$ without 4-vertices, e.g.\
$G=G^\Sigma$ (see section 3). By (4.28) we can then introduce an empty
tube connecting $\Sigma\times \{0\}$ and $\Sigma\times \{1\}$ thus
obtaining a handlebody with $2g$ handles.  Using (4.8) the graph on
its boundary can be reduced to one without 3-vertices and using in
addition Lemmas 3.4-5 and (4.9) the handles can be cut yielding a
3-ball with a graph on its boundary. The corresponding state sum can
then be evaluated and shown to equal the RHS of (5.3) by Lemma 4.3
{\it iii)}. $\hfill \Box$

We note here that, as a consequence of eqs. (4.24-25),
$|\vec{N}|^2$ on the RHS of (5.3) can be replaced by
 \be {\vec{N}}^2\equiv \sum_a (N^a)^2  \, ,\ee
where the symmetric matrix $N^a$ is defined by
\be{(N^a)}_{bc}=N^a_{bc}
, \;\;\; b,c\in I\, ,\ee
for each $a\in I$. Thus
\be dim V_{\Sigma}= (tr (\vec{N}^{2(g-1)}))^2 \, .\ee

Let now  $G_g$ be the graph depicted below

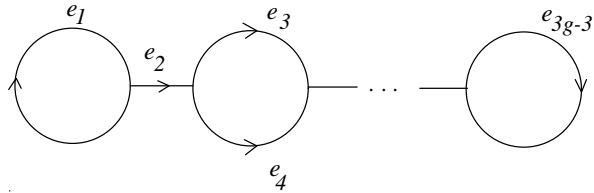
\begin{center}
\mbox{\epsfysize=2.5cm\epsffile{kanon.ps}}
\end{center}
with $g$ circles and coloured by $\underline{e}=(e_1, ..., e_{3g-3})$
for $g > 1$ and $\underline{e}=(e_1)$ for $g=1$.
Notice that the dimension of the vector space $V(\Sigma, G_g)$
associated to the graph $G_g$ embedded in a closed oriented
surface $\Sigma$ is given by
\be dim V(\Sigma,G_g)= (tr (\vec{N}^{2(g-1)}))^2 \, .\ee
as is easily seen by use of Lemma 4.3.

Let $M_{\Sigma^g}\subseteq R^3$ be a handlebody with $g$ handles and
$\partial M_{\Sigma^g}=\Sigma^g$, containing in its interiour two
disjoint (non-oriented) copies
$c^L_g$ and $c^R_g$ of $G_g$ as deformation retracts of $\Sigma^g$.
Thus, in particular, the circles in $c^L_g$ and $c^R_g$ are
non-contractible in $M_{\Sigma^g}$. We let ${\cal G}^L_g$  and ${\cal G}^R_g$
be two disjoint interiour graphs in $M_{\Sigma^g}$ whose cores are
 $c^L_g$ and $c^R_g$, respectively, and whose associated graphs
 $ G^L_g$  and $ G^R_g$  are also copies of $G_g$ on the boundaries of
tubular neighborhoods ${\cal T}^L$ and  ${\cal T}^R$, respectively,
which are equipped with $3g-3$ meridians for $g>1$ and one for $g=1$
 each such that all lines in
 $ G^L_g$, resp. $ G^R_g$, are lefthanded, resp. righthanded. Denote
the vector spaces associated to  $G^L_g\subseteq \partial {\cal T}^L$
and  $G^R_g\subseteq \partial {\cal T}^R$
by $V^L_g$ and $V^R_g$, respectively.

By introducing a canonical graph $G$ without 4-vertices on $\Sigma_g$ we
may consider the partition function associated to $M_{\Sigma_g}$ with interiour
graphs
${\cal G}^L_g$  and ${\cal G}^R_g$  and boundary graph $G$ as a linear
mapping from $(V^L_g\otimes V^R_g)^\ast$ into $V_\Sigma$.
It is in fact not hard to show that this mapping is an isomorphism. We
shall not need this general result in the following, but restrict our
attention to the case $g=1$.

In this case $G_g$ is a circle and $\Sigma_g $ is a torus $T$ embedded
in $R^3\subseteq S^3$ and $M_T$ is a solid torus, whose complement $\bar{M}_T$
in $S^3$ is also a solid torus with  $\partial \bar{M}_T=T^\ast$.
We thus consider copies $c^L$, $c^R$, resp.  $\bar{c}^L$, $\bar{c}^R$,
of the circle $S^1$ in $M_T$ and $\bar{M}_T$, respectively, and interiour
graphs ${\cal G}^L$  and ${\cal G}^R$, resp. ${\bar{\cal G}}^L$  and
${\bar{\cal G}}^R$,  whose associated tubular neighborhoods ${\cal T}^L$
and ${\cal T}^R$, resp. ${\bar{\cal T}}^L$  and
${\bar{\cal T}}^R$, are solid tori, and whose associated graphs $G^L$,
$G^R$, resp. $\bar{G}^L$, $\bar{G}^R$, are oriented circles, which we
assume have zero linking number with $c^L$, $c^R$,  $\bar{c}^L$,
$\bar{c}^R$, respectively. Moreover, $G^L$ and $\bar{G}^L$ are
lefthanded and $G^R$, $\bar{G}^R$ are righthanded and we note that
$c^L$ and $\bar{c}^L$, resp. $c^R$ and $\bar{c}^R$, are linked as
indicated in the following picture:

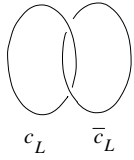
\begin{center}
\mbox{\epsfysize=2cm\epsffile{link.ps}}
\end{center}
 Since the graphs  $G^L$, $\bar{G}^L$,  $G^R$, $\bar{G}^R$
have no 3-vertices and only one line the vector spaces associated to
them may be identifyed with ${\bf C}^{|I|}$, where $|I|$ is the
cardinality of I, (cf. eq.(5.7)).

The partition function $Z^\prime_T$ (as defined in eq.(3.4))
associated to $M_T$ with interiour graphs
 ${\cal G}^L$  and ${\cal G}^R$ and a canonical graph $G$ without
4-vertices on $T$ thus yields a linear mapping
$$K:{\bf C}^{|I|}\otimes {\bf C}^{|I|}\to V_T$$
and similarly the partition function $\bar{Z}^\prime_T$ associated to
$\bar{M}_T$ with the interiour graphs
 ${\bar{\cal G}}^L$  and
${\bar{\cal G}}^R$ and boundary graph $G$ on $T^\ast$ yields a linear
mapping
$$L:V_T\to{\bf C}^{|I|}\otimes {\bf C}^{|I|} \, .$$

Gluing $M_T$ and $\bar{M}_T$ with interiour graphs along $T$ yields,
of course, $S^3$ with interiour graphs
 ${\cal{G}}^L$, ${\cal{G}}^R$,  ${\bar{\cal G}}^L$,
${\bar{\cal G}}^R$. Thus the mapping $LK:{\bf C}^{|I|}\otimes {\bf
C}^{|I|} \to {\bf C}^{|I|}\otimes {\bf C}^{|I|} \, $
is given by the martix
\be (LK)_{(e,e^\prime) (f, f^\prime)}=Z(S^3,  {\cal{G}}^L_e \cup
{\cal{G}}^R_f \cup {\bar{\cal G}}^L_{e^\prime}\cup {\bar{\cal G}}^R_{f^\prime}
) = \omega^2 Z(S^3,  {\cal{G}}^L_e \cup {\bar{\cal G}}^L_{e^\prime})
Z(S^3,{\cal G}^R_f\cup {\bar{\cal G}}^R_{f^\prime}
) \, ,\ee
where we have used 5) at the end of section 4 and Lemma 4.4 {\it i)}.
According to Lemma 4.4 {\it ii)- iii)} we have

\be  Z(S^3,  {\cal G}^L_e \cup {\bar{\cal G}}^L_{e^\prime})=\omega^{-1}
S_{e,e^\prime} \; ,\ee
\be Z(S^3,{\cal G}^R_f\cup {\bar{\cal G}}^R_{f^\prime})=
\omega^{-1}  S^{-1}_{f,f^\prime} \; ,\ee
where $S_{e, e^\prime}$ is defined by eq. (4.15).
Thus eq. (5.8) reads
\be LK= \omega^2( \omega^{-1} S\otimes \omega^{-1} S^{-1})\, ,\ee
and we conclude now from the fact that $dim V_T=dim ({\bf C}^{|I|}\otimes
{\bf C}^{|I|})  $, that $L$ and $K$ are isomorphisms and that
\be  K(S^{-1}\otimes S)L=1_{V_T} \, ,\ee
provided the assumptions in Thm. 5.1 are fulfilled.

 It was first observed by M. Karowski and R. Schrader [KS1]
   that
the remarkable property of the vector spaces $V_\Sigma$ being factorized
into  lefthanded and righthanded ones should lead to a rather simple proof
of the relation  between two different constructions of the
3-manifolds invariants. This is the content of the following theorem
where the context is assumed to be that of quantum deformations of a
semisimple Lie algebra at an even root of unity, in which case the
assumptions of Thm. 5.1 are known to be fulfilled and in addition
$|q_a|=1$ for all $a\in I$.

\begin{satz} Let $M$ be a closed, compact, oriented 3-manifold. Then
\be Z(M)=\tau (M) \,\tau (M^\ast)\, ,\ee
where $\tau (M)$ is an invariant defined by eq. (5.17) below, which in
fact equals the invariant introduced in [RT2], up to normalization.
\end{satz}
{\bf Proof:}
As is well known $M$ can be obtained (up to diffeomorphism) by surgery
along a framed link in $S^3$: Let ${\cal L}_1, ..., {\cal L}_n$ be the
components of ${\cal L}$, i.e.\ ${\cal L}_i$ is a framed circle, and
let ${\cal T}_i$ be a tubular neighborhood of ${\cal L}_i$ such that
$\cap_i {\cal T}_i=\emptyset$. We then obtain $M$ as
$$(S^3/(\cup_i{\cal T}_i)) \cup_{F_1} {\cal T}_1\cup_{F_2} ...
\cup_{F_i}{\cal T}_i ... \cup_{F_n}{\cal T}_n \, ,$$
i.e.\ by  first removing all ${\cal T}_i$ from $S^3$ and then gluing
them back with identification maps $F_i:\partial {\cal T}_i
\to \partial {\cal T}_i\subseteq \partial (S^3/\cup_i {\cal T}_i)$,
which are determined by the framing of ${\cal L}_i$. In other words if the
framing of ${\cal L}_i$ is given by the circle $L_i$ on $\partial {\cal T}_i$
with linking number $N$ w.r.t.\ the core of ${\cal T}_i$, then $F_i$ is
composed of an inversion (which interchanges the cycles in a canonical
homotopy basis for $\partial {\cal T}_i$) and an $N$-fold Dehn-twist.
Equivalently, $L_i\subseteq  \partial {\cal T}_i$
is identified with a meridian on $\partial {\cal T}_i \subseteq
\partial (S^3/\cup_i{\cal T}_i)$.

Inserting the identity operator in the form of (5.12) for each
$\partial {\cal T}_i$ a simple calculation yields

\be Z(M)=\sum_{\underline{e}, \underline{f},
\underline{e^\prime},\underline{f^\prime}} Z(S^3,
  ({\cal L}_{\underline{e}})_L\cup ( {\cal L}_{\underline{f}})_R)
 \prod^n_{i=1}   S^{-1}_{e_i e^\prime_i} \delta_{e^\prime_i 0}
   S_{f_i f^\prime_i} \delta_{f^\prime_i 0}
\;, \ee
 where the factors $ \delta_{e^\prime_i 0}\; \delta_{f^\prime_i 0}$
arise because the inversions make the $e^\prime_i$- and $f^\prime_i$-lines
traverse handles and hence their colours vanish by Lemma 4.2.

Using (4.18) and Lemma 4.4 {\it i)} in (5.14) we get

\begin{eqnarray}
Z(M) & = & ({\omega^2})^{-n+1} \sum_{\underline{e}} \omega^2_{\underline{e}}
Z(S^3, ({\cal L}_{\underline{e}})_L) \sum_{\underline{f}}
\omega^2_{\underline{f}}
Z(S^3, ({\cal L}_{\underline{f}})_R) \nonumber \\
 & = & (\omega^2)^{-n+1} Z(S^3, {\cal L}_L)
Z(S^3, {\cal L}_R)\; ,
\end{eqnarray}
where $ \omega^2_{\underline{a}}=\prod^n_{i=1}\omega^2_{e_i}$.
%The Reshetikhin-Turaev invariants [RT2] for two possible orientations of
%the manifold $M$, obtained by the surgery on  $S^3$ along $n$-components
%link $\cal L$, can
%be written in our notation as:
Eq. (5.15) may be rewritten as
\be Z(M)=\tau_L(M)\tau_R(M)\, ,\ee
if we define
\begin{eqnarray}
\tau_L (M) & = & \omega^{-n+1} (\Delta_L \omega^{-1})^{\sigma({\cal
L})}
Z(S^3, {\cal L}_L)\nonumber\\
 & = & \omega^{-n-1} (\Delta_L \omega^{-1})^{\sigma({\cal
L})}\sum_{\underline{e}}  \omega^2_{\underline{e}}
Z(G_L({\cal L})_{\underline{e}})\; ,\end{eqnarray}
%\be\tau (M^\ast)=\omega^{-n-1} (\Delta_R \omega^{-1})^{\sigma({\cal
%L})}\sum_{\underline{f}}  \omega^2_{\underline{f}} F (\Gamma ( {\cal
%G}^R, \underline{f})) \, ,\ee
and correspondingly $\tau_R$ by replacing $L$ by $R$ in (5.17),
where we have used Lemma 4.4 {\it ii)} and {\it iii)}. Here
$\sigma({\cal L})$ is the signature of a certain 4-manifold whose
boundary is $M$, $\Delta_L=\Delta$ is defined by eq. (4.22) and $\Delta_R$
is defined by replacing
 $q_i$ by   $q^{-1}_i=\bar{q}_i$ in eq. (4.22).
Since it is known (see [T, ch.2]) that
$$\Delta_L \Delta_R=\omega^2 \, $$
it follows that (5.15) and (5.16) are equivalent.

In section 6 we show that
\be Z(G_L({\cal L})_{\underline{e}})=F(\Gamma({\cal L},
\underline{e}))\;,\ee
where $F$ denotes the functor from the category of coloured ribbon
graphs into the category of representations of the quantum group under
consideration, which was introduced in [RT1]. Using the identity
(5.18) in (5.17) one obtains the expression for the invariant
$\tau(M)$ introduced in [RT2] (up to normalization) as given in
[T, ch.2]. Thus $\tau_L(M)=\tau(M)$. Since, moreover, it is known that
$\tau(M^\ast)={\tau(M)}^\ast$ and it is easy to verify that $\tau_R(M)={
\tau_L(M)}^\ast$ as a consequence of the properties of 6j-symbols, we
have proven (5.13).
$\hfill \Box$
\vspace*{0.3cm}

As a final topic in this section we shall briefly discuss how vector
spaces can be associated to surfaces with punctures and the
corresponding extension of Thm. 5.1.

Let $\Sigma$ be a closed, compact, oriented surface with a set
${\underline{p}}=(p_1, ...,p_n)$ of $n$ different, distinguished
points, and consider a
canonical graph $G$ on $\Sigma$ without 4-vertices, disjoint from
$\{p_1, ...,p_n\}$. Adjoining to $G$ lines $l_1, ..., l_n$  such that
$l_i$ connects $p_i$ to a point in $G$, different from the vertices in
$G$ and otherwise not intersecting $G$, we obtain a graph $G({\underline{p}}) $
on
$\Sigma$ whose vertices are all of order 2 or 3 except for $p_1, ...,
p_n$, which are of order 1. The dual graph to $G({\underline{p}}) $ on
$\Sigma$,
constructed in the standard way, then yields the proper triangulation
of $\Sigma$ except for the $n$ discs $D_1, ..., D_n$ containing $p_1,
..., p_n$, respectively, whose boundaries consist of a single link
each, namely the dual links to $l_1, ..., l_n$. Interpreting these
discs as holes in $\Sigma$, what remains is a proper triangulation of
$\Sigma/\cup_i D_i$, such that the boundary of each hole consists of a
single link (see Fig.8)

\begin{center}
\mbox{\epsfysize=4.2cm\epsffile{punct1.ps}}
\\Fig.8
\end{center}

We call graphs of the form $G({\underline{p}})$ canonical graphs for the
punctured
surface $\Sigma(p)$ and we shall assume  that the discs $D_1, ...,
D_n$ around $p_1, ..., p_n$, respectively, are fixed and only consider
graphs $G$ disjoint from $\cup_i D_i$. Moreover, we assume that $l_i$
intersects $\partial D_i$ at a (unique) fixed point $q_i$.

Fixing the colours of the boundary links in the triangulation, i.e.\
of the lines $l_1, ..., l_n$  to be $\underline{a}=(a_1, ..., a_n)$
we define, in analogy with $V(\Sigma, G)$, the vector space
$V_{\underline{a}}(\Sigma (\underline{p}), G(\underline{p}))$ to be the
direct sum over colourings of the remaining lines in $G(\underline{p})$
of the tensor product of spaces associated to the coloured 3-vertices
in  $G(\underline{p})$.

We next have to define a set of linear mappings between these spaces
for different canonical graphs possessing properties corresponding to
those of $P_{(\Sigma,G_1),(\Sigma, G_2)}$
discussed in section 3. For this purpose let $G_1$ and $G_2$ be two
canonical graphs on $\Sigma$ and consider    the graphs
 $G_1(\underline{p})\times \{0\}$ and  $G_2(\underline{p})\times \{1\}$
on $\partial (\Sigma\times [0,1])$. We construct the graph
$G_1(\underline{p})+ G_2(\underline{p})$
on $\partial ((\Sigma/\cup_i D_i)\times [0,1])$
by connecting the points $(q_i,0)$ in  $G_1(\underline{p})\times
\{0\}$ to $(q_i,1)$ in  $G_2(\underline{p})\times \{1\}$
by the $a$-coloured line $q_i\times [0,1]$, suitably oriented.
Finally, we introduce meridians $m_i$ on the tubes $T^i=\partial
D_i\times [0,1]$. It is then easy to see that
$(G_1(\underline{p})+ G_2(\underline{p}))\cup \{m_i\} $
is a canonical graph on
 $\partial ((\Sigma/\cup_i D_i)\times [0,1])$ and that the
corresponding vector space with $\underline{a}$ fixed is
$${V_{\underline{a}}(\Sigma (\underline{p}), G_1(\underline{p}))}^\ast \otimes
V_{\underline{a}}(\Sigma (\underline{p}), G_2(\underline{p}))\, .$$
The partition function
$$Z^\prime_{\underline{a}}(M(\{T^i_{m_i}\}),G_1(\underline{p})+
G_2(\underline{p})) $$
defined as in eq. (3.4) but with $\underline{a}$ fixed, may hence be
considered as a mapping
$$P_{G_1,G_2}(\underline{a}): V_{\underline{a}}(\Sigma (\underline{p}),
G_1(\underline{p})) \to
V_{\underline{a}}(\Sigma (\underline{p}), G_2(\underline{p}))\, .$$

Since this mapping depends on the chiralities of the
$\underline{a}$-lines w.r.t.\ the meridians we change notation and
denote by $\underline{a}=(a_1, ..., a_r)$, resp.  $\underline{b}=(b_1,
..., b_s)$, the colours of the left-, resp. right-, handed
lines and the correspondingly replace $\underline{a}$
everywhere by $\underline{a}, \underline{b}$.

It is a simple consequence of Thm. 3.1 and the projection property of
meridians that for any three canonical graphs $G_1,G_2,G_3$ without
4-vertices on $\Sigma$ we have
$$P_{G_2,G_3}(\underline{a},\underline{b}) P_{G_1,G_2}(\underline{a},
\underline{b})  = P_{G_1,G_3}(\underline{a},\underline{b})$$
and, in particular,
$P_{G}(\underline{a},\underline{b})=P_{G,G}(\underline{a},\underline{b}) $
is a projection. Hence we let $V^\prime_{\underline{a}, \underline{b}}(\Sigma
(\underline{p}), G(\underline{p}))\, $
be the support of this projection and conclude, as in section 3, that
these spaces for different $G$ may be consistently identified with a
space $ V_{\underline{a}, \underline{b}}(\Sigma (\underline{p}))$
by the mappings $ P_{G_1,G_2}(\underline{a},\underline{b})$
and that
\be dim  V_{\underline{a}, \underline{b}}(\Sigma (\underline{p}))
=Z(\Sigma\times S^1, {\cal L}^L_{\underline{a}}\cup \bar{{\cal
L}}^R_{\underline{b}}) \ee
where $\cal L$, resp. $\bar{\cal L}$, consists of $r$, resp. $s$,
framed circles of the form $p\times S^1$ with trivial framing and
${\cal L}^L_{\underline{a}}$, resp. $\bar{{\cal L}}^R_{\underline{b}}$,
denote the corresponding coloured left-,
resp. right-, handed links.

On the basis of (5.17) a slight extension of the proof of Thm. 5.1
yields the following generalization (see Thm. 7.4 in [KS]).

\begin{satz} Let $\Sigma (\underline{p})$ be a connected, closed,
compact, oriented surface of genus $g\ge 1$ with $n$ punctures $(p_1,
..., p_n)\equiv \underline{p}$, $r$ of which are lefthanded with
colours $\underline{a}=(a_1, ..., a_r)$ and $s$ righthanded with
colours $\underline{b}=(b_1, ..., b_s)$. Then
\be dim  V_{\underline{a}, \underline{b}}(\Sigma (\underline{p}))=
tr(N_{a_1}\ldots
N_{a_r}{|\vec{N}|}^{2(g-1)}) tr(N_{b_1}\ldots
N_{b_s}{|\vec{N}|}^{2(g-1)}) \; \ee
\end{satz}

We remark that in (5.18) we have assumed a certain common orientation
of the $\underline{a}$  and $\underline{b}$ coloured lines in the
definition of $V_{\underline{a}, \underline{b}}(\Sigma (\underline{p}))$.

 \section{Ribbon graphs and 6j-symbols}\label{qg}
\input epsf.sty
\newcommand{\alp}{\alpha
\kern-0.6em\lower1.8ex\hbox{$\tilde{\phantom{\alpha}}$}}
\newcommand{\alpb}{\alp
\kern-0.7em\lower2.2ex\hbox{$\tilde{\phantom{\alp}}$}}
The purpose of this section is partly to describe how the properties
of the 6j-symbols of a quantum group can be demonstrated by the use
of the  notion of ribbon graphs and partly to verify eq. (5.18).

As was proven in [A],
 $q$-deformations of the universal enveloping algebras  of complex
simple finite-dimensional Lie algebras $U_q {\sf g}$
   (quantum groups)  with
$q$-admissible modules  $\{V_i\}$ constructed in [TW] for
$q=\exp{(i\pi/l)}$ and  $l$ bigger then the
Coxeter number  of ${\sf g }$ give us
natural examples
of  modular Hopf algebras.
 (For the Lie algebras of type $ {\sf
g}_2$  $l$ must also be  odd and not divisible by 3.)
For the reader's convenience we recall some of the basic notions that
enter the definition of a modular Hopf algebra.

{\bf Definition [TW]:}
Let be $(A,R,v)$ a ribbon Hopf algebra  (see Def.3.3 in [RT1]) over ${\bf
C}$, where $A$ is a
quasi triangular Hopf algebra, $R$ a universal $R$-matrix and
 $v\in A$  a central element with  special properties (see eq. (6.3) below).
Assume  the following data are given:

{\it i)} a finite set $I$ with involution $i\mapsto i^\ast:I\to I$ and
a preferred element $0=0^\ast$,

{\it ii)} a set of $A$-modules $\{V_i\}$ labeled by $i\in I$, where
$V_0={\bf C}$ with the action of $A$ determined by the counit $A\to
{\bf C}$,

{\it iii)} a set of  $A$-linear isomorphisms
$$\{{ {\it w}}_i:(V_i)^\ast\to V_{i^\ast}, i\in I \},\;\;\;{ {\it w}}_0={\rm
id}_{\bf C}$$
The triple $(A,R,v)$ together with these data is called a modular Hopf
algebra if the following axioms (1-5) are satisfied:

1) The modules $\{V_i, i\in I\}$ are mutually non-isomorphic,
irreducible (i.e. do not contain proper non-trivial $A$-submodules),
have a finite  {\bf C}-dimension and all have nonsero
quantum dimension (see def.  below).

2) For each  $i\in I$ the homomorphism $${ {\it w}}^\ast_i\circ({\it
w}_{i^\ast} )^{-1}:V_i\to V^{\ast\ast}_i=V_i $$
is the multiplication by $g=uv^{-1}$.

3) For any $k\geq 2$ and for any sequence  $\Theta=(\lambda_1, \ldots,
\lambda_k)\in I^k$ there exists an $A$-linear decomposition:
$$ V_{\lambda_1}\otimes V_{\lambda_2}\otimes\ldots V_{\lambda_k}=
Z_\Theta \oplus \bigoplus_{\lambda\in I} (V_\lambda \otimes
\Omega^\lambda_\Theta)$$
where $\{\Omega^\lambda_\Theta\}$ are vector spaces over
  $\bf C$  and $Z_\Theta$ is a certain $A$-module satisfying the next
axiom (4).

4) For any $k\geq 2, \Theta\in I^k$ and any $A$-linear
homomorphism $f:Z_{\Theta}\to Z_{\Theta}$ the  $q$-trace of $f$
is equal to $0$, where  $q$-trace
of the  operator $f:V\to V$ is defined as the  trace of
an  operator $$x\mapsto uv^{-1}f(x):V\to V$$  and the quantum
dimension ${\rm dim}_q V$ equals
${\rm tr}_q({\rm id}_V)$, where $u\in A$ is such that that $ad_u$
equals the square of the antipode $S$ of $A$ (see also eq. (6.3) below).

5) Let $S_{ij}$ the $q$-trace  of the $A$-linear operator
$$ a\mapsto R_{21}R_{12}a:V_i \otimes V_j\to V_i \otimes V_j\, ,$$
Then the matrix $S_{ij}$ must be invertible.
\vspace*{0.3cm}

We shall henceforth use the standard notations for ribbon graphs and
let $F$ denote the functor introduced in [RT1] mentioned previously.

In order to prove eq. (5.18) it is enough to verify that $Z$ and $F$
agree on the graphs depicted in (4.3-6), i.e.\
that the equations analogous to (4.3-6) with $Z$ replaced by $F$ hold.
In fact, one uses repeatedly Lemma 3.2 and Lemma 4.2 {\it ii)-iii)}
(most effectively in the form of the Wigner-Eckart type relations
derived in [KS]) to decompose any planar graph $G$ into pieces of the
types in (4.3-6) and correspondingly obtains $Z(G)$ as a contraction of
a linear combination of tensor products of partition functions of the
pieces. Since the analogue of Lemma 4.2 also holds for the functor $F$
(see below) the claim follows.

The validity of the analogues of eqs. (4.3-6) for $F$ is, on the other
hand, essentially obvious by inspection. In fact eq. (4.3) holds for
$F$ since
 $\omega^2_x$  equals the quantum dimension of
$V_x, x\in I$, up to a sign, and eq. (4.4) for $F$
 can be obtained  by a suitable choice of dual bases $\{\alpha\}$ and
$\{\alpha^\ast\}$
in the mutually dual intertwiner  spaces
 $V^k_{ij}$ (from $V_i\otimes V_j$ to $V_k$) and
$V^{ij}_k$ (from $V_k$ to $V_i\otimes V_j$), so that
\be\label{loop} \mbox{\epsfysize=4cm\epsffile{qloop.ps}} \ee
which is the ribbon graph version of (4.7).

We note that the dual basis  $\{\alpha^\ast\}$ in
$ V^{ij}_k$ to  $\{\alpha\}$ in $V^k_{ij}$  can be constructed
 w.r.t.\  a natural bilinear pairing or, alternatively by exploiting
the natural inner product on $V^k_{ij}$
 (see [D]).
%$$(x,y)_G=(x,(\pi_{V_i}\otimes\pi_{V_j})(G)y)\, , $$
% where  $x,y\in V_i\underline{\otimes} V_j , \; G=R(q\otimes q)(\Delta
%q^{-1})\,$ and $q_i$  (for each $i\in I$) now is a specially root from
%$v_i$, which is defined  in [D92]. In this case
%$$ \alpha^\ast=(\pi_{V_i}\otimes\pi_{V_j})(G)\alpha^t\, ,$$
%where $t$ means  the transposition w.r.t.\ the scalar product.

Eq. (4.5) can be written in terms of ribbon graphs as
%\be \vspace*{0.2cm}\label{js}\vbox to 3cm{\special{psfile=sech.ps hscale=65
%%vscale=65
%hoffset=-75 voffset=-155}}\hspace*{3.cm} \doteq \hspace*{0.3cm}{\la
%%i&j&k\\l&m&n \ra }^{\alpha
%\beta^\ast \gamma^\ast \delta}\; , \ee
% \newpage
\be
\mbox{\parbox{5.1cm}{ \label{aa}\hbox{
\psfig{figure=sech.ps,height=5.2cm,width=5.1cm} } }\vspace*{0.2cm}
\parbox{4.2cm}{$\doteq\;\;\;{\la i&j&k\\l&m&n \ra }^{\alpha
\beta^\ast \gamma^\ast \delta}\; $}}
\ee
and the analogue of eq. (4.6) for $F$ is given below in (6.6).

We next proceed to show that the  6j-symbols defined by (6.2) have the
required
properties (see section 2). First we note that  due to the lemma
5.1 of [RT2]
  ribbon graphs can be considered as  two-side objects, where
the down sides of  the ribbons and annuli
have the dual colour and  opposite orientation compared  to the
up side.
We shall use  an additional operation  on the ribbons
and annuli --  so-called  half-twists, as proposed by [N]. The left
half-twist is illustrated in Figs.9 and 10.

\begin{center}
\mbox{\epsfysize=3.5cm\epsffile{twist1.ps}}
\\ Fig.9 \hspace*{2.7cm} Fig.10
\end{center}
The transformation in Fig.9  is given by the  action of the operator
 $\pi_{V_i}(\tau):V_i\to (V_{i^\ast})^\ast=V_i$,
and the one in Fig.10 by the action of
$ \pi_{V^\ast_i}(\tau)\colon V^\ast_i\to V_{i^\ast}=V^\ast_i $.
We use here the standard notation:
$$\pi_{V^\ast_i}(a)=[\pi_{V_i}(S(a))]^t, \; a\in A,$$
where $S(a)$ denotes the  antipode of $a$, $t$ the transposition w.r.t.\
the canonical pairing $V^\ast_i\otimes V_i\to {\bf C}$ and $\tau$ an
invertible element of the Hopf algebra, which satisfies the following
identities:
\be \tau^2=v^{-1}, \,u={\rm Ad}_\tau S(u), \,\varepsilon(\tau)=1,
\\ \Delta(\tau)=(\tau\otimes\tau)R,\; S(\tau)=\tau^{-1}u^{-1}.\ee
$\tau$ corresponds to the element ${\it w}^{-1}$ in the [RT1] notation.
The definition of the right half-twists is obtained from the
above   by replacing of  $\tau$  by $\tau^{-1}$.

It turns out that  in such a way  defined half-twists have a  nice geometrical
property, namely they can be pulled off from
 left to right through
all four annihilation and creation generators of the ribbon graph
category:

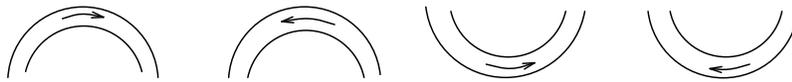
\begin{center}
\mbox{\epsfysize=1cm \epsffile{gen1.ps}}
\end{center}
This  follows directly from the properties of
$\tau$ for non self-dual  colours. In the case, when $i=i^\ast$,
there  exists an intertwiner $T:V_i\to
V^\ast_i$, and the above  mentioned property can be reduced to the claim:
 $T=T^t$. This means that the two isomorphisms
$V^\ast_i\otimes V_i \cong V_i\otimes V_i$ and
$V_i\otimes V^\ast_i \cong V_i\otimes V_i$  give us  the same invariant vector
$\xi\in V_i\otimes V_i$, i.e.\
\be\label{xi} Perm(\pi_{V_i}(g)\otimes 1) \xi = \xi\, .\ee
This was proven for quantum groups in  [DJN]. In the terminology of
 [T]  (\ref{xi})  distinguishes  unimodular and
modular Hopf algebras.

We may now construct   isomorphisms  $V^k_{ij}\simeq V^{j^\ast}_{k^\ast
i}\simeq  V^{i^\ast}_{j k^\ast}$ as follows (cf. also [DJN]). To each
$\alpha\in V^k_{ij}$ we associate
an $\tilde{\alpha}\in V^{j^\ast}_{k^\ast i}$ and an $\alp \in
V^{i^\ast}_{j k^\ast}$, such that
 \begin{center}
\mbox{\epsfysize=4cm\epsffile{alpha.ps}}
\\Fig.11
\end{center}
It is easy to see from the well known identities for ribbon graphs [RT1],
that $$\tilde{\tilde{\alpha}}=\alp\; ,
 \tilde{\tilde{\tilde{\alpha}}}=\alpha\; ,
\alpb=\tilde{\alpha}\,.$$

Due to the fact, that   all admissible modules
 $\{V_i\}$ of a modular Hopf algebra  have  real $q$-dimension
 there exist    isomorphisms $\alpha\to \alpha^G$ between $V^k_{ij}$
and $V^k_{ji}$ given by:
\begin{center}
\mbox{\epsfysize=3cm\epsffile{spiegel1.ps}}
\\Fig.12
\end{center}
where  $q_i=\sqrt{v^{-1}_i}$, $ v_i=\pi_{V_i}(v)$, $v_i=v_{i^\ast} $.
 The modular property of ribbon Hopf algebras
$R_{21}R_{12}(v^{-1}\otimes v^{-1})\Delta v=1$  implies
${\alpha}^{GG}=\alpha$.

On the other hand the   identity
$\Delta(\tau)=(\tau\otimes\tau)R$
allows us to construct an   isomorphism  between $V^k_{ij}$ and $V^{i^\ast
j^\ast}_{k^\ast}$:

\begin{center}
\mbox{\epsfysize=4cm\epsffile{dual.ps}}
\\Fig.13
\end{center}
 We denote  by  ${\cal V}^k_{ij}$ the   $\bf C$-module
which is obtained by identifying
 $V^k_{ij}, V^{j^\ast}_{k^\ast
i},  V^{i^\ast}_{j k^\ast}, V^k_{ji} $ and $V^{i^\ast
j^\ast}_{k^\ast}$ by the isomorphisms constructed above.
If we  choose  in (6.2)
 $\alpha\in {\cal V}^k_{ij},\,\delta\in {\cal
V}^m_{kl},\,
\gamma^\ast\in {\cal V}^{jl}_n,\, \beta^\ast\in
{\cal V}^{in}_m$ the  symmetry properties (2.7) follow  directly  from the
isotopies of the corresponding ribbon graphs [RT1].
  Additionally,  we can show  using the
isomorphisms in Fig.12 and Fig.13, that
\be \label{dual}
\mbox{\parbox{4.2cm}{
${\la i^\ast&j^\ast&k^\ast\\l^\ast&m^\ast&n^\ast \ra
}^{\alpha^\ast
\beta \gamma \delta^\ast} =$}\vspace*{0.2cm}
\parbox{4.9cm}{
\hbox{
\psfig{figure=stsech.ps,height=5.6cm,width=4.8cm}  }} .}
\ee

Due to the properties 3) and  4) of the  modular Hopf algebras
 for closed graphs we always have the equivalence
\be\label{gabel} \mbox {\epsfysize=4cm\epsffile{qgabel.ps}}\ee
which is the ribbon graph version of (4.8).
%From (\ref{loop}) and (\ref{gabel}) follows immediately: $\sum_k\omega^2_k
%N^k_{ij} =\omega^2_i \omega^2_j $. The invertibility of $\omega$ and
%$\Delta$ (lemma 4.3) were proved in [T] for  modular Hopf algebras.
%This is  $\omega^2_{k^\ast}=
%(\omega^2_k)^\ast$ and for quantum groups  $\omega^2_k=\omega^2_{k^\ast}$.

Taking  (\ref{dual}), (\ref{loop}) and (\ref{gabel}) into
consideration one can quite
 easily  prove
the  \OR, the \BE\  and the Racah identities  for  6j-symbols defined
by (6.2).  Applying isomorphisms given by Fig.12 to each 3-vertex
in (6.2) and using isotopies of ribbon graphs one obtains (2.8).
The last condition  (2.12) is a consequence of the normalization (\ref{loop}).
In fact, to preserve (\ref{loop}) for $k=0$
we must introduce the factor $\omega^{-1}_i$
after  removing the  0-ribbon from  $\alpha\in V^0_{i i^\ast}$.

Finally, one can analogously prove that
\be
\mbox{\parbox{5.1cm}{ \label{aa}\hbox{
\psfig{figure=racah.ps,height=5.2cm,width=5.1cm} } }\vspace*{0.2cm}
\parbox{4.2cm}{$=\frac{q_A q_B}{q_C q_D}\la y&A&D\\x&B&C \ra \; $}}
\ee
which is the ribbon graph version of eq. (2.17).

\section{Conclusion}
We have in this paper developed calculational techniques applicable to
a large class of 3-dimensional state sum models of topological quantum
field theories, and extending those of [KS]. As a byproduct we have
calculated the dimensions of the state spaces associated to surfaces
and obtained a relatively simple proof of the relation between these
models and those introduced in [RT1,2], and hence to continuum
Chern-Simons theory with an arbitrary compact semisimple gauge group
[Wi2]. Our arguments have been mainly based on simple geometrical
considerations which we believe can be generalized rather
straightforwardly to higher dimensions. Indeed, higher dimensional
models of Turaev-Viro type have been suggested recently in the
literature and it would be an obvious task to introduce observables
into those models in the form of ''higher dimensional graphs''
 by extending our construction in section 3.
In order to develop such a construction into an effective
calculational tool it would be crucial to have at disposal analogues
of the Racah identities. We leave these issues for future investigation.

\vspace*{1cm}

{\Large \bf Acknowledgements}
\vspace*{1cm}

The first author would like to thank M. Karowski and R. Schrader
for many fruitful discussions, and the second author acknowledges the
kind hospitality extended to him at the Yukawa Institute for
Theoretical Physics, Kyoto, in the fall of 1992, where part of this
work was carried out.

\end{document}